\documentclass[longauth]{aa} 

\usepackage{graphicx}
\usepackage{lscape}

\usepackage{txfonts}
\usepackage{mathabx}
\usepackage[normalem]{ulem}
\usepackage{color}
\usepackage[dvipsnames]{xcolor}
\usepackage{soul}

\def \MJ{M$_{\mathrm{Jup}}$}
\def \MN{M$_{\mathrm{Nep}}$}
\def \ME{M$_{\Earth}$}
\def \RJ{R$_{\mathrm{Jup}}$}
\def \RN{R$_{\mathrm{Nep}}$}
\def \RE{R$_{\Earth}$}
\def \RS{R$_{\odot}$}
\def \msol{M$\mathrm{_\odot}$}

\def \kms{km\,s$^{-1}$}
\def \ms{m\,s$^{-1}$}
\def \1s{$1\,\sigma$}

\def \t0{T$_0$}

\def \toi17{TOI~1736}
\def \toipl17{TOI~1736~b}

\begin{document} 

    \title{TOI-3568~b: a super-Neptune in the sub-Jovian desert}
    \titlerunning{Detection and characterization of TOI-3568~b}   
    
   \author{
    E. Martioli\inst{\ref{lna},\ref{iap}} 
    \and R. P. Petrucci \inst{\ref{unc},\ref{conicet}} 
    \and E. Jofr\'e  \inst{\ref{unc},\ref{conicet}} 
    \and G. H\'ebrard \inst{\ref{iap},\ref{ohp}} 
    \and L. Ghezzi \inst{\ref{ufrj}} 
    \and Y. G\'omez Maqueo Chew  \inst{\ref{unam}} 
    \and R. F. D\'iaz \inst{\ref{icas}} 
    \and H. D. Perottoni \inst{\ref{lna}} 
    \and L. H. Garcia \inst{\ref{unc}} 
    \and D. Rapetti  \inst{\ref{nasa-ames},\ref{riacs}}  
    \and A. Lecavelier des Etangs \inst{\ref{iap}} 
    \and L. de\,Almeida \inst{\ref{lna}} 
    \and L. Arnold \inst{\ref{cfht}} 
    \and \'E. Artigau \inst{\ref{montreal}} 
    \and R. Basant \inst{\ref{uc}} 
    \and J. L. Bean \inst{\ref{uc}} 
    \and A. Bieryla \inst{\ref{cfa}} 
    \and I. Boisse \inst{\ref{lam}} 
    \and X. Bonfils \inst{\ref{ipag}}  
    \and M. Brady \inst{\ref{uc}} 
    \and C. Cadieux \inst{\ref{montreal}} 
    \and A. Carmona \inst{\ref{ipag}} 
    \and N. J. Cook \inst{\ref{montreal}} 
    \and X. Delfosse \inst{\ref{ipag}} 
    \and J.-F. Donati \inst{\ref{irap}} 
    \and R. Doyon \inst{\ref{montreal}} 
    \and E. Furlan \inst{\ref{caltech}} 
    \and S. B. Howell \inst{\ref{nasa-ames}} 
    \and J. M. Jenkins \inst{\ref{nasa-ames}} 
    \and D. Kasper \inst{\ref{uc}} 
    \and F. Kiefer \inst{\ref{lesia},\ref{iap}} 
    \and D. W. Latham  \inst{\ref{cfa}} 
    \and A. M. Levine \inst{\ref{mit}} 
    \and D. Lorenzo-Oliveira \inst{\ref{lna}} 
    \and R. Luque \inst{\ref{uc}} 
    \and K. K. McLeod \inst{\ref{wellesley}} 
    \and J. Melendez \inst{\ref{iagusp}} 
    \and C. Moutou \inst{\ref{irap}} 
    \and Y. Netto \inst{\ref{iagusp}} 
    \and T. A. Pritchard \inst{\ref{nasa-goddard}}  
    \and P. Rowden \inst{\ref{royal-ast-soci}} 
    \and A. Seifahrt \inst{\ref{uc}} 
    \and G. Stef\'ansson \inst{\ref{univamsterdam}} 
    \and J. St{\"u}rmer \inst{\ref{heidelberg}} 
    \and J D. Twicken \inst{\ref{seti},\ref{nasa-ames}}  
    }

    \institute{
    Laborat\'{o}rio Nacional de Astrof\'{i}sica, Rua Estados Unidos 154, 37504-364, Itajub\'{a} - MG, Brazil, \email{martioli@iap.fr} \label{lna}
    \and Institut d'Astrophysique de Paris, CNRS, UMR 7095, Sorbonne Universit\'{e}, 98 bis bd Arago, 75014 Paris, France \label{iap}
    \and  Universidad Nacional de C\'ordoba - Observatorio Astron\'{o}mico de C\'{o}rdoba, Laprida 854, X5000BGR, C\'ordoba, Argentina \label{unc}
    \and Consejo Nacional de Investigaciones Cient\'{i}ficas y T\'{e}cnicas (CONICET), Godoy Cruz 2290, CABA, CPC 1425FQB, Argentina \label{conicet}
    \and Observatoire de Haute Provence, St Michel l'Observatoire, France \label{ohp}       
    \and Universidade Federal do Rio de Janeiro, Observat\'orio do Valongo, Ladeira do Pedro Ant\^onio, 43, Rio de Janeiro, RJ 20080-090, Brazil \label{ufrj}
    \and Instituto de Astronom\'ia, Universidad Nacional Aut\'onoma de M\'exico, Ciudad Universitaria, Ciudad de M\'exico, 04510, M\'exico \label{unam}
    \and International Center for Advanced Studies (ICAS) and ICIFI (CONICET), ECyT-UNSAM, Campus Miguelete, 25 de Mayo y Francia, (1650) Buenos Aires, Argentina. \label{icas}
    \and NASA Ames Research Center, Moffett Field, CA  94035, USA \label{nasa-ames}
    \and Research Institute for Advanced Computer Science, Universities Space Research Association, Washington, DC 20024, USA   \label{riacs}
    \and Canada-France-Hawaii Telescope, CNRS, 96743 Kamuela, Hawaii, USA \label{cfht}
    \and Universit\'{e} de Montr\'{e}al, D\'{e}partement de Physique, IREX, Montr\'{e}al, QC H3C 3J7, Canada \label{montreal}
    \and Department of Astronomy \& Astrophysics, University of Chicago, Chicago, IL, USA \label{uc}
    \and Center for Astrophysics \textbar\ Harvard \& Smithsonian, 60 Garden Street, Cambridge, MA 02138, USA \label{cfa}
    \and Aix Marseille Univ, CNRS, CNES, LAM, 38 rue Frédéric Joliot-Curie, 13388 Marseille, France  \label{lam} 
    \and Universit\'{e} Grenoble Alpes, CNRS, IPAG, 414 rue de la Piscine, 38400 St-Martin d'Hères, France \label{ipag}
    \and Universit\'e de Toulouse, CNRS, IRAP, 14 avenue Belin, 31400 Toulouse, France \label{irap}
    \and NASA Exoplanet Science Institute, Caltech IPAC, 1200 E. California Blvd., Pasadena, CA 91125, USA \label{caltech}       
    \and LESIA, Observatoire de Paris, Universit\'e PSL, CNRS, Sorbonne Universit\'e, Universit\'e Paris Cit\'e, 5 place Jules Janssen, 92195 Meudon, France \label{lesia}    
    \and Department of Physics and Kavli Institute for Astrophysics and Space Research, Massachusetts Institute of Technology, Cambridge, MA 02139, USA \label{mit}  
    \and Department of Astronomy, Wellesley College, Wellesley, MA 02481, USA \label{wellesley}
    \and Universidade de São Paulo, Instituto de Astronomia, Geofísica e Ciências Atmosféricas (IAG), Departamento de Astronomia, Rua do Matão 1226, Cidade Universitária, 05508-900, SP, Brazil \label{iagusp}
    \and NASA Goddard Space Flight Center, 8800 Greenbelt Road, Greenbelt, MD 20771, USA \label{nasa-goddard}
    \and Royal Astronomical Society, Burlington House, Piccadilly, London W1J 0BQ, United Kingdom \label{royal-ast-soci}
    \and Anton Pannekoek Institute for Astronomy, University of Amsterdam, Science Park 904, 1098 XH Amsterdam, The Netherlands \label{univamsterdam}    
    \and Landessternwarte, Zentrum f{\"u}r Astronomie der Universität Heidelberg, K{\"o}nigstuhl 12, D-69117 Heidelberg, Germany \label{heidelberg}
    \and SETI Institute, Mountain View, CA  94043, USA \label{seti}
   }
    
   \date{Received xxxx ; accepted xxxx }

  \abstract {
    The sub-Jovian desert is a region in the mass-period and radius-period parameter space, typically encompassing short-period ranges between super-Earths and hot Jupiters, that exhibits an intrinsic dearth of planets. This scarcity is likely shaped by photoevaporation caused by the stellar irradiation received by giant planets that have migrated inward.  We report the detection and characterization of TOI-3568~b, a transiting super-Neptune with a mass of $26.4\pm1.0$~\ME, a radius of $5.30\pm0.27$~\RE, a bulk density of $0.98\pm0.15$~g\,cm$^{-3}$, and an orbital period of 4.417965~(5)~d situated in the vicinity of the sub-Jovian desert.  This planet orbiting a K dwarf star with solar metallicity, was identified photometrically by the Transiting Exoplanet Survey Satellite (TESS). It was characterized as a planet by our high-precision radial velocity monitoring program using MAROON-X at Gemini North, supplemented by additional observations from the SPICE large program with SPIRou at CFHT.  We performed a Bayesian MCMC joint analysis of the TESS and ground-based photometry, MAROON-X and SPIRou radial velocities, to measure the orbit, radius, and mass of the planet, as well as a detailed analysis of the high-resolution flux and polarimetric spectra to determine the physical parameters and elemental abundances of the host star. Our results reveal TOI-3568\,b as a hot super-Neptune, rich in hydrogen and helium with a core of heavier elements with a mass between 10 and 25~\ME. We analyzed the photoevaporation status of TOI-3568~b and found that it experiences one of the highest EUV luminosities among planets with a mass M$_{p}<2$~M$_{\rm Nep}$, yet it has an evaporation lifetime exceeding 5~Gyr. Positioned in the transition between two significant populations of exoplanets on the mass-period and energy diagrams, this planet presents an opportunity to test theories concerning the origin of the sub-Jovian desert.
  }

   
   \maketitle
%
\section{Introduction}

Exoplanets with sizes ranging between Jupiters and super-Earths and orbiting their host stars at short periods $<5$~days are notably rare, giving rise to what is termed the sub-Jovian or Neptunian desert \citep{SzaboKiss2011,Mazeh2016}. The prevailing hypotheses to explain this intrinsic deficit in the planet population involve high-eccentricity migration and stellar irradiation \citep{Owen2018}. This combination is believed to cause planetary inflation or the erosion of primary atmospheres of exoplanets through photoevaporation, imposing constraints on planet mass and radius depending on their distance from the star.

Recent confirmations of several exoplanets residing within the sub-Jovian desert \citep[see][]{Szabo2023,Kalman2023,Frame2023} showed that this region is not completely barren. Instead, it represents a less probable condition for planets to exist. The sparse presence of planets within the sub-Jovian desert suggests that the formation and evolutionary trajectories of these exoplanets may have taken a unique path, diverging from those observed in more densely populated regions of parameter space. Ongoing surveys optimized for redder stars, such as the Transiting Exoplanet Survey Satellite \citep[TESS,][]{tess_paper}, are detecting an increasing number of planets in the sub-Jovian desert, particularly around smaller stars \citep[e.g.,][]{Eigmuller2017,West2019,Eigmuller2019,Murgas2021,Mori2022}.  Increasing this sample is important to test physical mechanisms involved in planet formation across a range of stellar parameters, including spectral types, masses, metallicities, and galactic populations.

Here, we present the discovery and characterization of TOI-3568~b, a hot super-Neptune orbiting a K-dwarf star located in the sub-Jovian desert, within a transitional region between the populations of hot Jupiters and short-period super-Earths, where planets are notably scarce. This discovery emerged from a program aimed at identifying and characterizing planetary systems within the thick-disk galactic population. Stars from distinct galactic populations exhibit differences in kinematics and chemical composition, potentially leading to variations in the frequency of giant planets as suggested by the planet-metallicity correlation \citep{Fischer2005,Johnson2010,Buchhave2014,Wang2015,Ghezzi2018}. We selected TOI-3568.01 as a planet candidate from the TESS Object of Interest (TOI) catalog \citep{Guerrero2021}, based on its high thick disk to thin disk membership probability (TD/D = 3.58) determined through kinematic classification \citep{Carrillo2020}. However, our analysis showed that the nature of this system is more consistent with what is typically associated with the thin disk.

This paper is organized as follows. In Section \ref{sec:observations} we present the observations and data reduction; in Sections \ref{sec:star} and \ref{sec:planetcharacterization} we present the characterizations of the star and the planet, respectively; in Section \ref{sec:discussions} we discuss the characteristics of this new planetary system in the context of the population of exoplanets; and we conclude in Section \ref{sec:conclusions}.
 
\section{Observations and data reduction}
\label{sec:observations}

\subsection{TESS photometry}
\label{sec:tessphotometry}

TOI-3568 (TIC 160390955) was first observed by the TESS in sector 15 with a cadence of 30 minutes. A planet candidate with a 4.42~d period was identified in the MIT Quick Look Pipeline (QLP) faint star transit search \citep{Kunimoto2022}. An alert for TOI-3568.01 was issued by the TESS Science Office on 23 June 2021. We performed a pre-analysis of the 30-minute cadence TESS Full-Frame Image (FFI) data using the community Python package \textsc{Lightkurve}\footnote{\url{https://docs.lightkurve.org}} \citep{lightkurvecollaboration} to obtain the photometric time series. We inspected the light curve data and analyzed the transits, where we employed the methods described in \cite{Martioli2020} and obtained a well-constrained model for the planetary parameters. Thus, we concluded that the events observed by TESS were probably planetary. We submitted TOI-3568 to the TESS Director’s Discretionary Targets (DDT 062, PI: E. Martioli), where we were able to include it for observations in sectors 55 and 56 in the 2-min cadence mode.  A subsequent search of the 2-min data from sectors 55 and 56 by the TESS Science Processing Operations Center (SPOC) pipeline detected the planetary signature of TOI-3568~b. The difference image centroiding test \citep{Twicken2018} located the host star to lie within $1.1\pm2.9$~arcsec of the source of the transits. Table \ref{tab:tessobservations} shows the log of TESS observations of TOI-3568. 

\begin{table}
\centering
\footnotesize
\caption{Log of TESS observations of TOI-3568.}
\label{tab:tessobservations}
\begin{tabular}{ccccc}
\hline
Sector & TSTART & TSTOP & Duration & Cadence  \\
 & (UTC) & (UTC) & (d) & (min) \\
\hline
15 & 2019-08-15 & 2019-09-11 & 26.0  & 30 \\
55 & 2022-08-05 & 2022-09-01 & 27.2  & 2 \\
56 & 2022-09-02 & 2022-09-30 & 27.9 & 2 \\
\hline
\end{tabular}
\end{table} 

\begin{figure}
   \centering
       \includegraphics[width=1.\hsize]{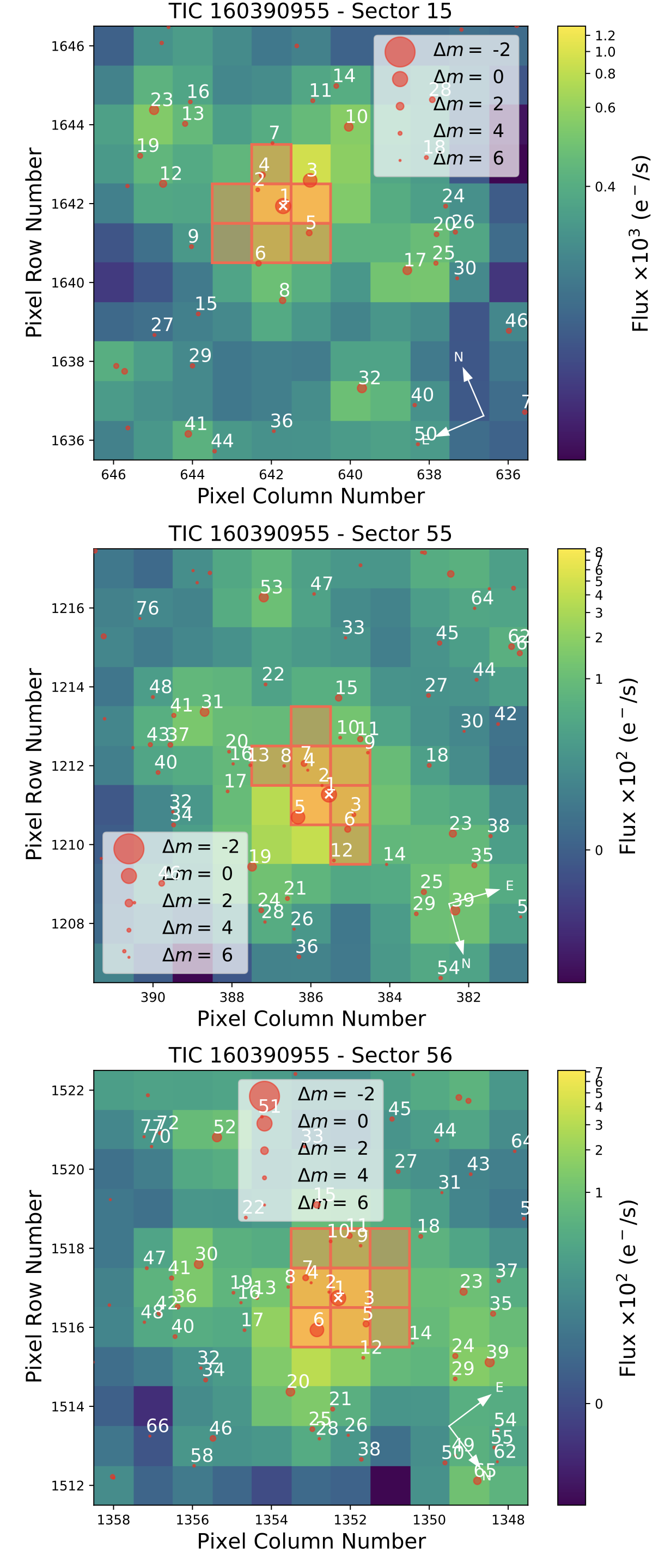}
      \caption{Target pixel files for TOI-3568 observed in TESS sectors 15 (top panel), 55 (middle panel), and 56 (bottom panel). The red circles represent the Gaia DR3 sources, with TOI-3568 marked by a cross and labeled as index 1. Symbol sizes correlate with the Gaia DR3 source's brightness compared to TOI-3568's brightness. The red-shaded squares indicate the pixels used in aperture photometry. The three images were produced with \textsc{tpfplotter} \citep{aller2020}. 
      }
        \label{fig:toi3568_tpf_gaia}
  \end{figure}

For the observations of TOI-3568 obtained in sectors 55 and 56, we first used the Presearch Data Conditioning (PDC) flux time series \citep{Smith2012,Stumpe2012,Stumpe2014} processed by the SPOC pipeline at NASA Ames Research Center \citep{jenkinsSPOC2016} obtained from the TESS data products available in the Mikulski Archive for Space Telescopes (MAST)\footnote{\url{mast.stsci.edu}}. Figure \ref{fig:toi3568_tpf_gaia} shows the TOI-3568's target pixel files (TPF) for sectors 15, 55, and 56. It highlights the pixels used in the aperture for photometry and marks the positions of sources in the field from Gaia DR3's catalog \citep{gaia2016, gaia2022, katz2023}.

We calculated our final TESS photometry using an optimized systematic error correction algorithm, following the methodology of Rapetti et al.~(in prep.). We use an adaptation of the Pixel Level Decorrelation~\citep[hereafter PLD;][]{Deming_2009, Luger_2016, Luger_2018} technique implemented in the \textsc{PLDCorrector} class of \textsc{Lightkurve}. This method employs (i) a spline polynomial fit to describe stellar variability; (ii) Principal Component Analysis (PCA) eigenmodes to model the background light; and (iii) the PLD technique to account for pointing and mechanical effects.  

To account for the background as described in (ii), we wish to begin with calibrated pixels that are not corrected for the background. We thus add the background flux estimated by SPOC into the calibrated and background-removed pixel values in the original TPF before the correction. Since PLD might preserve the mean of the uncorrected light curve after the regression, to recover the true mean flux level of the corrected light curve we apply a flux level adjustment. We adjust the corrected flux by subtracting a constant level calculated as the third dimmest median pixel flux value times the number of pixels. This is inspired by the scalar background bias method applied in the SPOC pipeline. We then adjust the flux for crowding by non-target stars and for the fraction of the target star flux captured in the photometric aperture using the methods of and the crowding and flux fraction values provided by the SPOC pipeline. We also use the flux fraction to scale the flux errors, but not the crowding since its effect on the flux errors is negligible.

\begin{figure}
   \centering
       \includegraphics[width=1.\hsize]{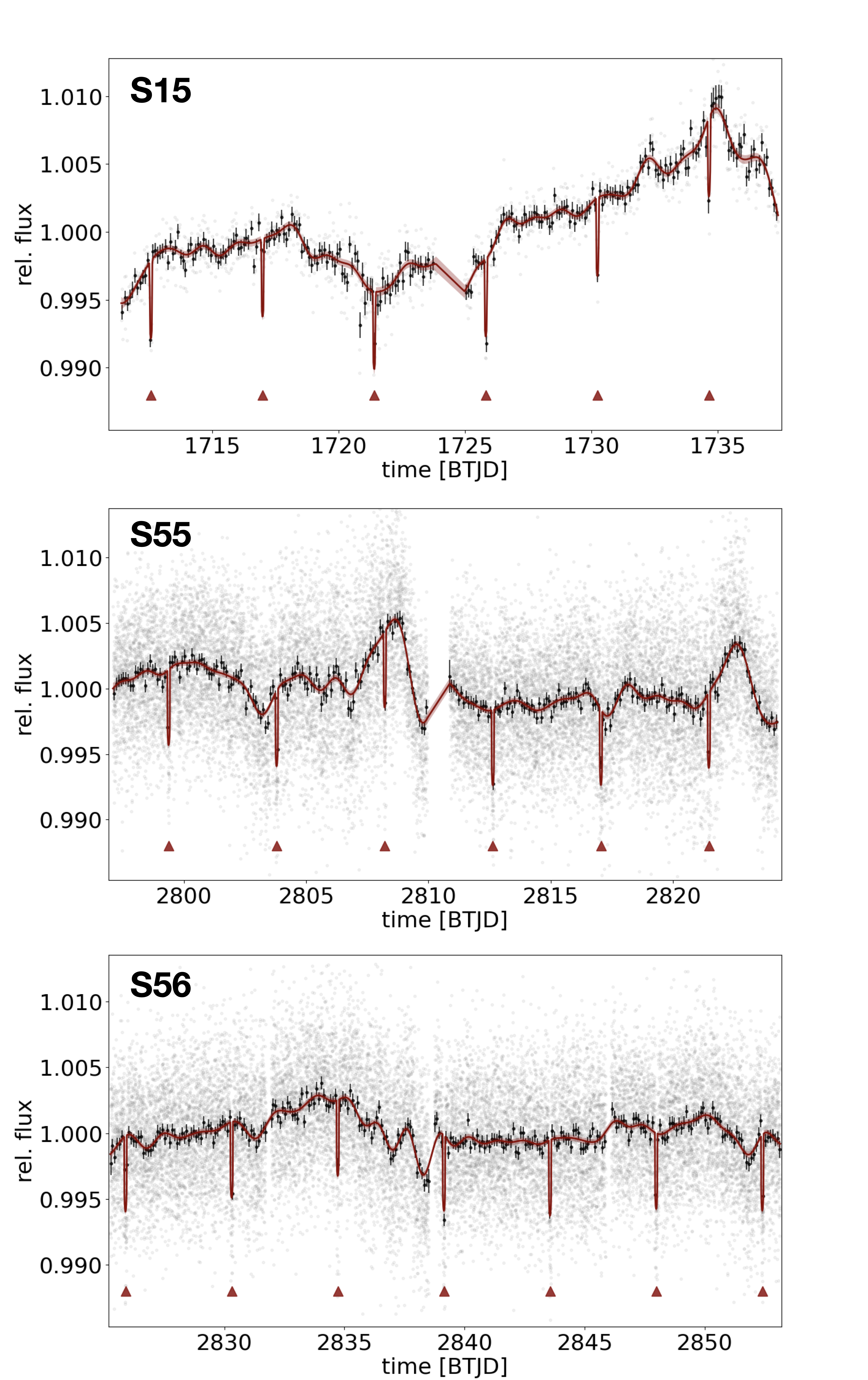}
      \caption{TESS light curve of TOI-3568. The gray points show the TESS photometry data normalized by the median flux of each sector. The black points show the same data binned by the weighted average with bin sizes of 0.1~d. The red line shows the baseline GP model that was fitted to the binned data multiplied by the best-fit transit model for TOI-3568~b. The red triangles show the central time of TOI-3568\,b transits.
      }
        \label{fig:toi3568_tess_lc}
  \end{figure}

Before applying the PLD corrector, we add the background flux and errors estimated by the SPOC pipeline back onto the Simple Aperture Photometry (SAP) light curve. Flux level, fraction and crowding adjustments are applied to the corrected light curves. To automatically optimize the selection of parameter values for the corrector, we evaluate the resulting light curve using the Savitsky-Golay Combined Differential Photometric Precision (sgCDPP) proxy algorithm \citep{Gilliland_2011,Van_Cleve_2016} implemented in \textsc{Lightkurve}, for durations of 30, 60, 120, 160, and 200 minutes. For a grid of corrector parameter values (for further details on the parameters and the grid, see Rapetti et al.~(in prep.)), we calculate the harmonic mean (HM) of these quantities and select the corrected light curve that minimizes the HM. Using this analysis, we were able to recover segments of the data that were initially excluded by the SPOC pipeline due to scattered light. The TESS light curves are illustrated in Figure \ref{fig:toi3568_tess_lc} along with the results of our model, as outlined in Section \ref{sec:analysisofTESSdata}.

\subsection{Ground-based photometry}
\label{sec:ground-basedphotometry}

Ground-based time-series photometry was collected as part of the TESS Follow-up Observing Program Sub Group 1 \footnote{\url{https://tess.mit.edu/followup/}} \citep{collins:2019} which uses a customized TESS version of the \textsc{TAPIR} software package \citep{Jensen2013} for observation planning. The data reduction and aperture photometry were performed using AstroImageJ \citep{Collins:2017} and the light curves are available on ExoFOP\footnote{https://exofop.ipac.caltech.edu}.

We observed a full transit event of TOI-3568~b on 2022-12-13 using the 0.7~m telescope at the Wellesley College Whitin Observatory \footnote{\url{https://www.wellesley.edu/whitin-observatory}} (WCWO) in MA, USA. Images were taken in an Sloan Digital Sky Survey (SDSS)$-r'$ filter using 30\,s exposures, and photometry was extracted using a circular aperture with a radius of 2.7~arcsec.  The aperture was small enough to exclude the light from the two nearest GAIA DR3 stars with projected separations of 3.7~arcsec and 5.7~arcsec, the latter of which is bright enough to have caused the TESS event if it had been an eclipsing binary. These uncontaminated $r'$ data are used along with the TESS light curves in the joint analysis in Section \ref{sec:planetcharacterization}.

\subsection{High contrast imaging}
\label{sec:highcontrastimaging}

High-angular resolution observations of candidate systems hosting transiting exoplanets can aid in identifying blended sources within sub-arcsecond scales. These sources might produce a false positive transit signal, particularly if the source is an eclipsing binary. The EXOFOP-TESS website\footnote{https://exofop.ipac.caltech.edu} reports five high contrast imaging observations, spanning from the optical range at 562\,nm to the near-infrared (NIR) at 2.2~$\mu$m, as summarized in Table \ref{tab:highcontrastobservations}. These observations indicate the absence of a close-in companion with sufficient brightness to generate a false positive signal.

\begin{figure}
   \centering
   \includegraphics[width=0.9\hsize]{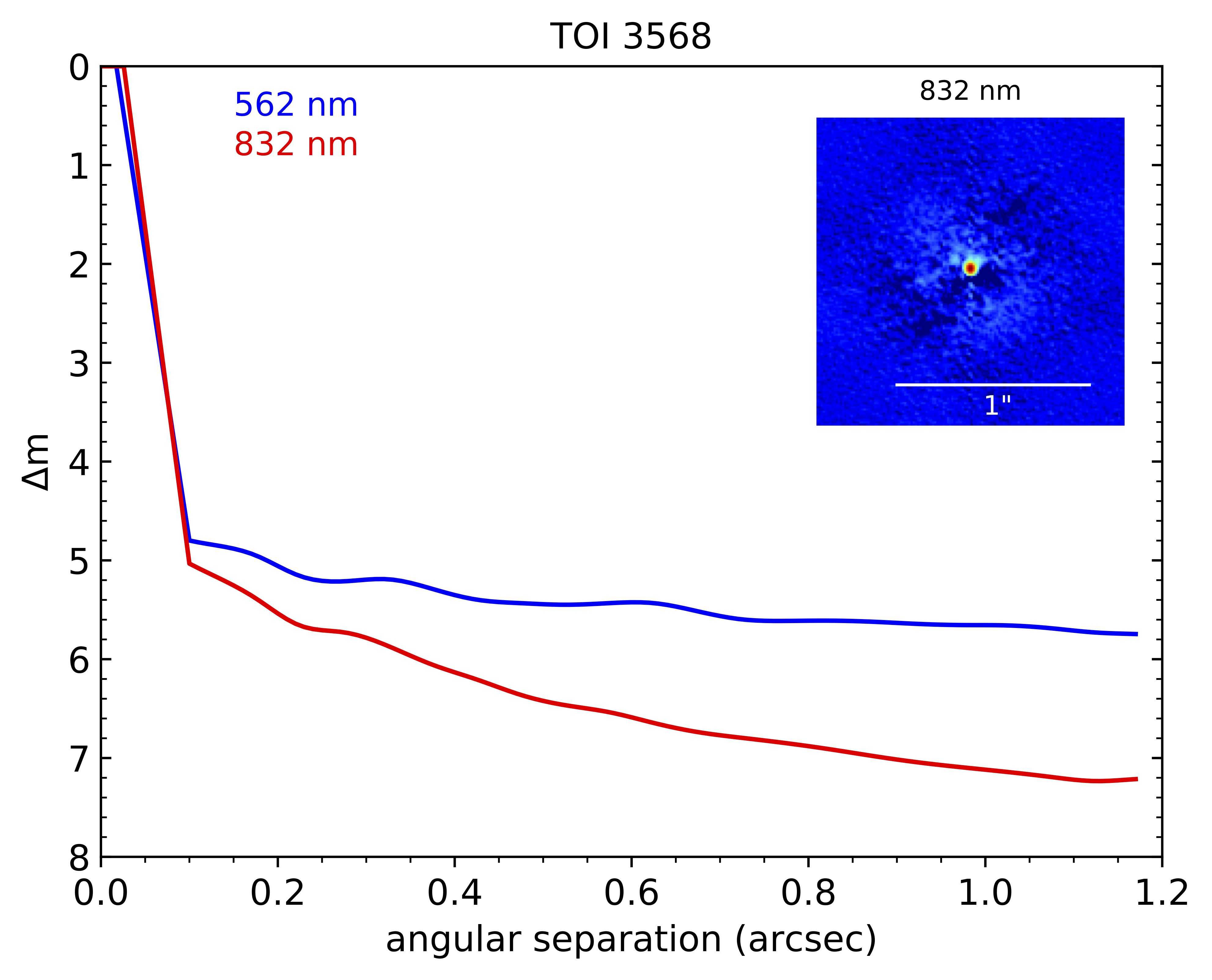}
      \caption{Contrast sensitivity for TOI-3568 as a function of angular separation at 562 nm (blue line) and at 832 nm (red line) obtained from the `Alopeke/Gemini-N speckle imaging observations. The inset panel displays the reconstructed speckle image at 832 nm.  
       }
        \label{fig:gemini_AO_constrast_TOI-3568}
  \end{figure}

Figure \ref{fig:gemini_AO_constrast_TOI-3568} shows the 5-$\sigma$ sensitivity curves derived from observations made with the `Alopeke dual-channel speckle imaging instrument on Gemini-N (PI: Howell). These observations were obtained with a pixel scale of 0.01~arcsec per pixel and a full width at half maximum (FWHM) resolution of 0.02~arcsec. The data were processed with the speckle pipeline \citep{Howell2011}. `Alopeke performed simultaneous speckle imaging at 562~(54)~nm and 832~(40)~nm. The results from these observations effectively eliminate the possibility of any companion with a contrast of $\Delta$mag$\,<6.42$ at 0.5~arcsec separation at 832~nm. This wavelength range is particularly pertinent to the photometry and spectroscopic observations presented in this paper, as it overlaps with the spectral sensitivity of TESS and MAROON-X.

\begin{table}
\centering
\tiny
\caption{Summary of high-contrast imaging observations of TOI-3568.}
\label{tab:highcontrastobservations}
\begin{tabular}{cccccc}
\hline
Observation & Telescope & Instrum. & $\Delta$mag  at &  Spectral   & Imaging \\
date (UTC) & site &  & 0.5~arcsec& band & technique \\
\hline
2022-09-13 & Gemini-N &  `Alopeke & 5.44  & 562~nm  & Speckle \\
2022-09-13 & Gemini-N &  `Alopeke & 6.42  & 832~nm  & Speckle \\
2021-08-28 & Keck2 &  NIRC2 & 6.77  & K-band   & AO \\
2021-08-24 & Palomar &  PHARO & 6.67  & Br-$\gamma$  & AO \\
2021-08-24 & Palomar &  PHARO & 6.88  & H-cont  & AO \\
\hline
\end{tabular}
\end{table} 

\subsection{MAROON-X spectroscopy}
\label{sec:maroonx}

MAROON-X is a high-resolution spectrograph ($\lambda / \Delta \lambda \sim 85000 $) operating in the optical range (500-920 nm) and installed on the 8.1-m Gemini North telescope atop Maunakea, Hawaii \citep{Seifahrt2018}. The spectrograph is fiber-fed, highly stabilized, and bench-mounted. It was designed to achieve sub-\ms\ radial velocity (RV) precision.

We obtained 34 spectra of TOI-3568 under program ID GN-2022A-Q-207/-Q-113 (PI: R. Petrucci) with an exposure time of 720~s and using MAROON-X in its single mode of operation. These spectra were collected over ten different nights spanning from 2022-04-08 to 2022-07-26. The strategy of obtaining $\sim3$ spectra per visit was precautionary, to mitigate the impact of potential outliers. On average, these observations yielded a peak signal-to-noise ratio (S/N) per spectral element of $51\pm11$ in the blue arm and $70\pm16$ in the red arm.

The MAROON-X raw data have been reduced using the standard procedure implemented in the instrument Python3 pipeline \citep{Seifahrt2020}. This procedure involved bias and background subtraction, order tracing and the extraction of one-dimensional wavelength-calibrated spectra. Wavelength solutions and instrumental drift corrections were based on the simultaneous calibration data of a stabilized Fabry–Pérot etalon \citep{Sturmer2017}, which allows order-by-order drift corrections at the sub–\ms\ level. The flux-weighted midpoint of each observation was used to calculate the barycentric corrections.

We analyzed the spectra using the SpEctrum Radial Velocity AnaLyser (SERVAL) pipeline \citep{Zechmeister2018}, which employs the template-matching algorithm to extract precise relative RVs. The blue and red channels of MAROON-X are reduced separately, producing independent RVs, as presented in Table \ref{tab:maroonx_rvs}.  The blue channel RVs show a root mean square (RMS) dispersion of $9.7$\ms and a median error of $1.9$\ms, whereas the red channel shows an RMS dispersion of $9.5$\ms and a median error of $3.5$\ms.

We co-added all individual MAROON-X observations shifted to the same stellar reference frame to obtain a master spectrum with high S/N. Our spectroscopic analysis in Section \ref{sec:star} uses this master spectrum.  

In addition, we employ a reference solar spectrum obtained from observations of sunlight reflected by the asteroid Vesta on the night of 2022-04-27 under program ID GN-2022A-Q-227 (PI: Y. Netto). These observations were carried out adopting the same MAROON-X setup (S/N $\sim$ 400 at 600 nm) to ensure precision in determining the stellar parameters and chemical abundances of TOI-3568 through a differential analysis.

\begin{table}
\centering
\small
\caption{
    MAROON-X relative radial velocities for the red and blue channels.
}
\label{tab:maroonx_rvs}
\begin{tabular}{ccccc}
\hline
 Time & $\Delta$RV & $\sigma_{\rm RV}$  & $\Delta$RV& $\sigma_{\rm RV}$ \\
 BJD &  \ms & \ms & \ms & \ms \\
 &  \multicolumn{2}{c}{Red channel} & \multicolumn{2}{c}{Blue channel} \\
\hline
2459678.102859 & -0.8 & 3.6 & -3.3 & 2.1 \\
2459678.112155 & 8.0 & 3.8 & 0.5 & 2.2 \\
2459678.121780 & 3.2 & 4.0 & -2.1 & 2.3 \\
2459724.061299 & -0.9 & 1.9 & 2.6 & 1.1 \\
2459724.076411 & -3.3 & 1.8 & 0.2 & 1.0 \\
2459732.095294 & 7.8 & 2.2 & 10.8 & 1.3 \\
2459732.110072 & 7.0 & 2.2 & 10.6 & 1.2 \\
2459769.067102 & -8.6 & 4.0 & -11.3 & 2.3 \\
2459769.076541 & -11.9 & 3.9 & -13.5 & 2.2 \\
2459769.086043 & -11.5 & 3.9 & -14.7 & 2.2 \\
2459773.033820 & 0.3 & 3.3 & -3.4 & 1.8 \\
2459773.043345 & -1.7 & 3.3 & -0.7 & 1.9 \\
2459773.052853 & -3.6 & 3.4 & -5.0 & 1.9 \\
2459773.105035 & -11.1 & 3.3 & -3.4 & 1.9 \\
2459773.114508 & -2.8 & 3.3 & -1.6 & 1.8 \\
2459773.124058 & -2.6 & 3.2 & -3.5 & 1.8 \\
2459775.984434 & 3.2 & 2.8 & 12.7 & 1.5 \\
2459775.994051 & 16.1 & 2.7 & 10.2 & 1.5 \\
2459776.003556 & 12.3 & 2.7 & 11.7 & 1.5 \\
2459776.066268 & 14.5 & 4.0 & 11.8 & 2.3 \\
2459776.075921 & 9.0 & 3.5 & 13.3 & 2.0 \\
2459776.085302 & 11.5 & 3.6 & 13.6 & 2.0 \\
2459778.850346 & -7.8 & 3.5 & -12.3 & 2.0 \\
2459778.859572 & -6.3 & 3.2 & -8.1 & 1.8 \\
2459778.868798 & -0.9 & 3.8 & -10.7 & 2.2 \\
2459781.106447 & 9.1 & 3.7 & 5.9 & 2.1 \\
2459781.116534 & 10.6 & 4.0 & 9.1 & 2.3 \\
2459781.126025 & 13.3 & 4.8 & 11.1 & 2.8 \\
2459783.065784 & -11.4 & 2.7 & -14.3 & 1.4 \\
2459783.075369 & -15.4 & 3.1 & -14.5 & 1.7 \\
2459783.084914 & -6.7 & 3.3 & -11.6 & 1.8 \\
2459787.098883 & -14.8 & 3.8 & -10.2 & 2.1 \\
2459787.108504 & -13.1 & 3.5 & -11.8 & 2.0 \\
2459787.117842 & -16.2 & 3.5 & -10.4 & 2.0 \\
\hline
\end{tabular}
\end{table} 

\subsection{SPIRou spectro-polarimetry}
\label{sec:spirou}

TOI-3568 was observed by the SpectroPolarimètre Infra-Rouge (SPIRou)\footnote{ \url{http://spirou.irap.omp.eu} and \url{https://www.cfht.hawaii.edu/Instruments/SPIRou/}} under the large program SPIRou Legacy Survey - Consolidation \& Enhancement (SPICE\footnote{\url{http://spirou.irap.omp.eu/Observations/The-SPIRou-Legacy-Survey}}; PI: Jean-Fran\c{c}ois Donati) on nights spanning from 2022-11-14 to 2022-11-21. SPIRou is a stabilized high-resolution near infrared (NIR) spectropolarimeter \citep{donati2020} mounted on the 3.6~m Canada-France-Hawaii Telescope (CFHT) atop Maunakea, Hawaii. It is designed for high-precision velocimetry to detect and characterize exoplanets and it provides a full coverage of the NIR spectrum from 950~nm to 2500~nm at a spectral resolving power of $\lambda / \Delta \lambda \sim 70000 $.

We observed TOI-3568 with SPIRou/CFHT at five different epochs, where we obtained a total of 20 spectra with an individual exposure time of 900~s. These observations were carried out in the circular polarization mode (Stokes~V), where each set of four exposures provides a polarimetric spectrum. The peak S/N per spectral element varied between 40 and 65, with a median of 59. The air mass of the observations ranged from 1.1 to 1.4 and the Barycentric Earth Radial Velocity (BERV) ranged from -20.0 to 20.7\,\kms. 

The SPIRou data have been reduced using the APERO pipeline v.0.7.284 \citep{Cook2022}, which produces 1D optimally extracted fluxes that underwent detector gain and artifact corrections,  wavelength calibration, blaze correction, and correction for telluric atmospheric absorption. Additionally, the pipeline computed polarimetric Stokes V and null spectra.  

The flux spectra have been analyzed using the line-by-line (LBL) method of \cite{Artigau2022}, wherein a high-S/N template spectrum of HD~189733 observed by SPIRou was used as a reference to obtain the RVs. The Table \ref{tab:spirou_rvs+blong} shows the SPIRou RVs. These RVs show an RMS of 12.8~\ms, and a median error of 12.4~\ms.

\begin{table}
\centering
\caption{
    SPIRou relative radial velocities and the $B_\ell$ data.
}
\label{tab:spirou_rvs+blong}
\begin{tabular}{ccccc}
\hline
Time & $\Delta$RV & $\sigma_{\rm RV}$ & $B_\ell$ & $\sigma_{B\ell}$  \\
BJD &  \ms & \ms & G & G \\
\hline
2459898.728015 & -12.5 & 11.3 & 16 & 13 \\
2459899.784369 & 8.4 & 12.6 & -22 & 31 \\
2459901.756303 & -20.0 & 11.0 & 18 & 17 \\
2459903.770162 & 14.6 & 12.4 & 17 & 22 \\
2459904.785654 & 0.0 & 18.6 & 1 & 32 \\
\hline
\end{tabular}
\end{table} 

\section{Stellar characterization} 
\label{sec:star}
We carried out a study to derive the host star properties as will be detailed in the next sections. Table \ref{tab:stellarparams} presents a summary of the stellar parameters of TOI-3568.

\begin{table*}
\centering
\caption{Summary of the stellar parameters for TOI-3568.}
\label{tab:stellarparams}
\begin{tabular}{lcc}
\hline
Parameter & Value & Ref. \\
\hline
ID (Gaia DR3)  & 1950477552581609984 & 1 \\
ID (TIC) & 160390955 & 2 \\
ID (2MASS)  & J21302453+3444226 &  3 \\
RA (hh:mm:ss.ss)  & 21:30:24.56943612 & 1 \\
Dec (dd:mm:ss.ss) & +34:44:22.48569168 & 1 \\
epoch (ICRS)  & J2000, epoch 2015.5 & 1 \\
galactic Longitude, $l$ (deg) & 82.69465 & 1 \\
galactic Latitude, $b$ (deg) & -11.98994 & 1 \\
proper motion in RA, $\mu_{\alpha}$ (mas\,yr$^{-1}$)  & $26.370\pm0.008$ & 1 \\
proper motion in Dec, $\mu_{\delta}$ (mas\,yr$^{-1}$) & $-7.854\pm0.010$ & 1 \\
parallax, $p$ (mas) & $5.056\pm0.012$ & 1 \\
distance (pc)  & $197.8\pm0.5$ & 1 \\
radial velocity (\kms)  & $-97.8\pm0.7$ & 1 \\
B (mag) &  $13.962\pm0.04$  & 2 \\
V (mag) &  $12.879\pm0.057$  & 2 \\
TESS T (mag)  & $12.0729\pm0.006$ & 2 \\
GAIA G (mag)  & $12.6870\pm0.0002$  & 1 \\
2MASS J (mag) & $11.189\pm0.021$  & 3 \\
2MASS H (mag) & $10.658\pm0.020$ & 3 \\
2MASS K (mag) & $10.581\pm0.011$ & 3 \\
WISE 1 (mag)  & $10.515\pm0.022$ & 4 \\
WISE 2 (mag) & $10.574\pm0.02$ & 4 \\
WISE 3 (mag) & $10.467\pm0.073$ & 4 \\
WISE 4 (mag) & $9.172$ & 4 \\
effective temperature, $T_{\rm eff}$ (K) & $4969\pm45$ & this work \\
surface gravity, $\log g$ (dex) & $4.63\pm0.08$ & this work \\
metallicity, $[{\rm Fe}/{\rm H}]$ (dex) &  $-0.01\pm0.02$ & this work \\
microturbulence velocity, $\nu$ (km/s) & $0.77\pm0.13$ & this work \\
star mass, $M_{\star}$ (\msol) &  $0.78\pm0.02$ & this work \\
star radius, $R_{\star}$ (\RS) & $0.719\pm0.013$ & this work \\
star density, $\rho_{\star}$ (g\,cm$^{-3}$) & $2.95\pm0.18$ & this work \\
luminosity, $\log{L_{\star}/L_{\odot}}$ &  $0.33\pm0.02$ & this work \\
projected rotational velocity, $v_{\rm rot}\sin{i_{\star}}$ (\kms) & $1.4\pm0.4$ & this work \\
age (Gyr) & $6.1\pm3.7$ & this work \\
age-[Y/Mg] (Gyr) & $7.6\pm1.2$ & this work \\
longitudinal magnetic field, $B_\ell$ (G) & $-2\pm15$ & this work \\
alpha-to-iron abundance ratio, $[{\rm \alpha}/{\rm Fe}]$ (dex) & $0.06\pm0.02$ & this work \\
\hline
\end{tabular}
\tablebib{
(1) \cite{GaiaEDR3Vizier};
(2) \cite{Stassun2018,Stassun2019}, sourced from the {\tt EXOFOP-TESS} website\footnote{https://exofop.ipac.caltech.edu};
(3) \cite{Cutri2003};
(4) \cite{Wright2010}.
}
\end{table*}

\subsection{Atmospheric parameters}
\label{sec:fundamentalparameters}

The fundamental atmospheric parameters (T$_{\mathrm{eff}}$, $\log g$, [Fe/H], and $v_{t}$) of TOI-3568 were determined by imposing a strictly line-by-line differential spectroscopic equilibrium of neutral and singly-ionized iron lines relative to the Sun \citep[e.g.,][]{Jofre2021}. To perform this process automatically, we employed the  \texttt{q2} program\footnote{The \texttt{q2} code is available at \url{https://github.com/astroChasqui/q2}} \citep{Ramirez2014}. The iron line list, as well as the atomic parameters, namely the excitation potential (EP) and oscillator strengths ($\log{gf}$), are the same as in \citet{Jofre2021}, and the equivalent widths (EWs) of the MAROON-X spectra of TOI-3568 and the Sun (reflected sunlight from Vesta) were manually measured by fitting Gaussian profiles using the \textsc{splot} task in \textsc{IRAF}\footnote{IRAF is distributed by the National Optical Astronomy Observatories, which are operated by the Association of Universities for Research in Astronomy, Inc., under cooperative agreement with the National Science Foundation.}. The solar values were kept fixed at (T$_{\mathrm{eff}}$, $\log g$, [Fe/H], $v_{t}$) = (5777~K, 4.44~dex, 0.0~dex, 1.0~\kms). The resulting stellar parameters are T$_{\mathrm{eff}}$=$4969\pm45$~K, $\log{g}=4.63\pm0.08$~dex, ${\rm [Fe/H]}=-0.01\pm0.02$~dex, and $v_{t}=0.77\pm0.13$~\kms. 

As an independent check, we derived fundamental parameters also using the excitation and ionization equilibria of Fe I and Fe II lines technique but without using a strictly line-by-line differential analysis \citep[e.g.,][]{Ghezzi2018, Ghezzi2021, Jofre2020}. We measured the EWs automatically with the \textsc{ARES} code \citep[e.g.,][]{Sousa2015} and the pipeline from \cite{Ghezzi2018} that uses the 2017 version of the \textsc{MOOG} code\footnote{https://www.as.utexas.edu/~chris/moog.html} \citep{Sneden1973} and a line list with solar log \textit{gf} values derived using the same setup (see \citealt{Ghezzi2018} for further details).
In excellent agreement with the parameters obtained with the differential technique we obtained: T$_{\mathrm{eff}}=4897\pm47$~K, $\log{g}=4.52\pm0.11$~dex, ${\rm [Fe/H]}=0.02\pm0.02$~dex, $v_{t}=0.52\pm0.16$~\kms.

We determined the projected star rotation velocity ($v_{\rm rot}\sin{i_{\star}}$) based on the spectral synthesis of relatively isolated iron lines using the code \textsc{iSpec} \citep{BlancoCuaresma2019} and following the procedure of \cite{Carlberg2012}. Adopting the calibration of \cite{Takeda2017} to determine a macroturbulence velocity of 1.06~\kms, we find  $v_{\rm rot}\sin{i_{\star}} = 1.4\pm0.4$~\kms. However, considering the  resolving power of MAROON-X (R = 85,000), we adopt an upper limit of 2 ~\kms.


\subsection{Mass, radius, and age} 
\label{sec:mass_radius_age}

We derived stellar mass, radius, and age using Yonsei-Yale (YY) stellar isochrones \citep{Yi2001, Demarque2004}, as described in \citet{Jofre2021}. This was accomplished via the \texttt{q2} pipeline, using as input the spectroscopic T$_{\rm eff}$ and [Fe/H] obtained from the differential method, Gaia DR3 parallax, and $V$-mag (corrected for extinction\footnote{Visual extinction (Av) is computed as a function of the stellar distance and the galactic coordinates (l, b) by interpolating in the tables given by \citet{Arenou1992} using Frédéric Arenou’s online calculator (\url{https://wwwhip.obspm.fr/cgi-bin/afm}).}). We obtained an age of $6.1\pm3.7$~Gyr, a mass of $0.780\pm0.021$~\msol, and a radius of $0.720\pm0.013$~\RS. 
This code also provides the trigonometric gravity, which allows us to perform a consistency check on the spectroscopic $\log g$ values. Here, \texttt{q2} yields $\log g = 4.62 \pm 0.02$ dex, which is in excellent agreement with the estimates found from the spectroscopic equilibrium.

As a consistency check, we also employed the 1.3 version of the \textsc{PARAM} web interface\footnote{\url{http://stev.oapd.inaf.it/cgi-bin/param_1.3}} that performs a Bayesian estimation of stellar parameters \citep{daSilva2006, miglio2013} based on PARSEC isochrones \citep{bressan2012}. As input, we employed the same parameters as those used above in \texttt{q2}. In good agreement, within the errors, with our estimations from \texttt{q2}, \textsc{PARAM} returned an age of $4.2\pm3.7$~Gyr, a mass of $0.75\pm0.02$~\msol, a radius of $0.69\pm0.01$~\RS, and $\log{g} = 4.60 \pm 0.02$ dex.

\subsection{Longitudinal magnetic field}
\label{sec:blong}

\begin{figure}
\centering
\includegraphics[width=1.0\hsize]{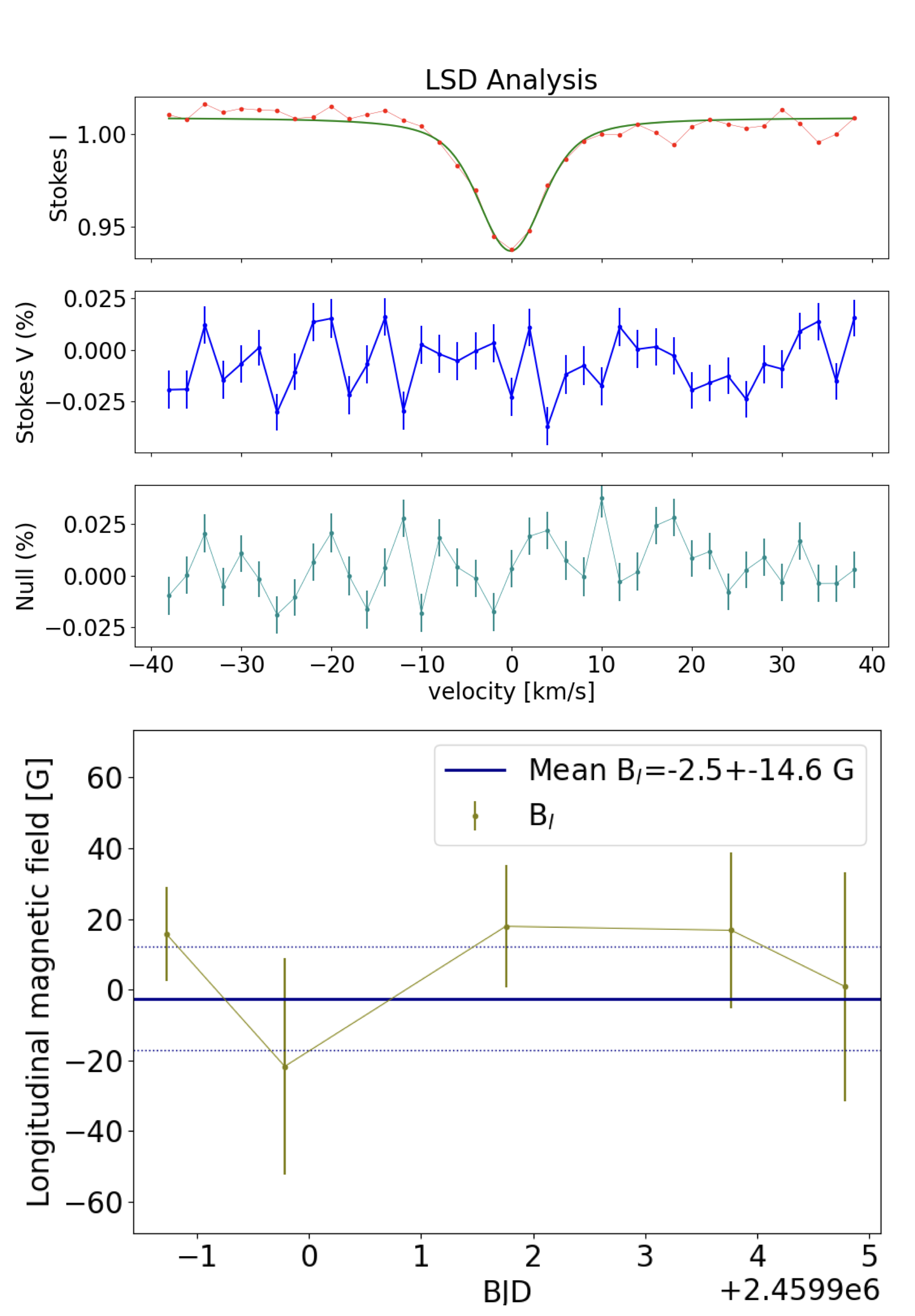}
\caption{ LSD analysis and longitudinal magnetic field from SPIRou spectropolarimetric data. The top panel shows the Stokes I (red points) with a Voigt profile model fit (green line), Stokes V (blue points), and null (dark cyan points) LSD mean profiles. The absence of the characteristic `S'-shape in the Stokes V profile indicates a non-detection of a Zeeman signature. In the bottom panel, the longitudinal magnetic field (green points) measured across the five epochs observed by SPIRou displays an average value of $\overline{B}_\ell=-2.5\pm14.6$~G, which is also consistent with a null value.}
\label{fig:spirou-polarimetry}
\end{figure}

We performed a least squares deconvolution (LSD) analysis and computed the longitudinal magnetic field ($B_\ell$) on individual SPIRou spectra for Stokes~I, Stokes~V, and null polarizations, using the methodologies introduced by \cite{donati97} and implemented by \cite{Martioli2020}. The resulting mean LSD profiles are depicted in the top panels of Figure \ref{fig:spirou-polarimetry}, while the time series of $B_\ell$ is illustrated in the bottom panel. Table \ref{tab:spirou_rvs+blong} shows the values of $B_\ell$. The Stokes~V profile is featureless, indicating the absence of a Zeeman signature for this star. The average longitudinal magnetic field, $\overline{B}_\ell=-2.5\pm14.6$~G, is consistent with a null detection of a magnetic field in TOI-3568, suggesting its magnetic inactivity.

\subsection{Activity}
\label{sec:activity}

To investigate activity in TOI-3568, we measured three spectral index proxies for chromospheric activity in the MAROON-X spectra: the Ca infrared triplet (Ca IRT), the H-$\alpha$ (available in both the blue and red channels), and the Na D1 and D2 doublet. We found no significant correlations between these indices and RVs, nor with the FWHM in either the blue or red channels, indicating that the RV variations are likely not caused by stellar activity. We measured the median values and standard deviations of the CaIRT and NaD indices as follows: CaIRT${1} = 0.446\pm0.004$, CaIRT${2} = 0.332\pm0.004$, CaIRT${3} = 0.447\pm0.005$, NaD${1} = 0.175\pm0.004$, and NaD$_{2} = 0.206\pm0.002$.  For H-$\alpha$, we measured $0.281 \pm 0.005$ in the blue channel and $0.283 \pm 0.007$ in the red channel, resulting in a mean value of $0.282 \pm 0.004$. This represents a variation in H-$\alpha$ of 1.4\% over a baseline of 100 days, indicating that TOI-3568 appears to have low levels of chromospheric activity.

We also looked for signs of stellar variability in the 2-min TESS PLD data from sectors 55 and 56. To do so, we carried out the analysis of the light curve with the tools in the \textsc{lightkurve} package. As a first step, based on the transit parameters obtained in our analysis (see Section \ref{sec:planetcharacterization}), we removed all the points falling within the transits of TOI-3568~b. Then, we ran two algorithms on the resulting light curve: the Lomb-Scargle periodogram \citep[LS;][]{scargle1982} and, an additional independent method, the Auto-Correlation Function \citep[ACF;][]{mcquillan2013}. After applying the criteria described in \citet{petrucci2024} to assess if a detected signal is real, we determined that no significant peak indicating periodic variability was found and, hence, there is no evidence for rotational modulation in the data. This may be a consequence of a small spot coverage on the stellar surface, or due to the existence of a rotation period longer than $\sim14$~d, which is likely undetected in the TESS data. Additionally, no flare candidate was detected on the 2-min cadence light curve by the \textsc{Altaipony} code \citep{davenport2016, ilin2021}, optimized to search for sporadic events. A careful by-eye inspection of the TESS photometry confirms these results.

The absence of signs of variability suggests that TOI-3568 is an inactive star. This aligns with the mature age value obtained in this study (see Table \ref{tab:stellarparams}) and with the low levels of chromospheric and magnetic activity measured from our spectropolarimetric data.

\subsection{Chemical composition} 
\label{sec:abundances}


We measured line-by-line differential abundances relative to solar ([X/H]) abundances of 20 elements other than iron, including C, O, Na, Mg, Al, Si, S, K, Ca, Sc, Ti, V, Cr, Mn, Co, Ni, Cu, Zn, Y, and Ba. This was achieved through EW measurements and by employing the curve-of-growth approach with the \textsc{MOOG} program (\textsc{abfind} driver) using the \texttt{q2} code. The EWs were manually measured using the \textsc{splot} task in \textsc{IRAF} and the adopted line list and atomic parameters were taken from \citet{Jofre2021}. Hyperfine splitting was taken into account for Sc, V, Mn, Co, Cu Y, and Ba. The O abundance was computed from the 7771-5~{\AA} IR triplet, adopting the non-LTE corrections by \citet{Ramirez2007}. 

Similar to the previous section, as an independent check, we also performed a non-strictly differential analysis to determine elemental abundances for TOI-3568 following \cite{Schuler2015} and \cite{Teske2015}.
The EWs were automatically measured for elements with more than three spectral lines using the code ARES \citep{Sousa2015} and an updated line list based on solar log \textit{gf} values that will be described in a future paper \footnote{The line list can be made available upon request directly from the authors if needed before the referred publication is released.} that investigates possible correlations between planetary properties and the chemical abundances of their stellar hosts. For elements with three or less lines, we manually measured the EWs with the \textsc{splot} task in \textsc{IRAF}.
The abundances were calculated using the \textsc{MOOG} code and \textsc{abfind} and \textsc{blends} drivers for the elements without and with hyperfine splitting, respectively. The \textsc{blends} driver was also used to determine the abundance from the [O I] 6300 \AA\ line in order to take the contamination from Ni into account. The non-LTE corrections for the O I 7771-5~{\AA}\ IR triplet were taken from \cite{Amarsi2019}.  

\begin{figure}
\centering
\includegraphics[width=1.0\hsize]{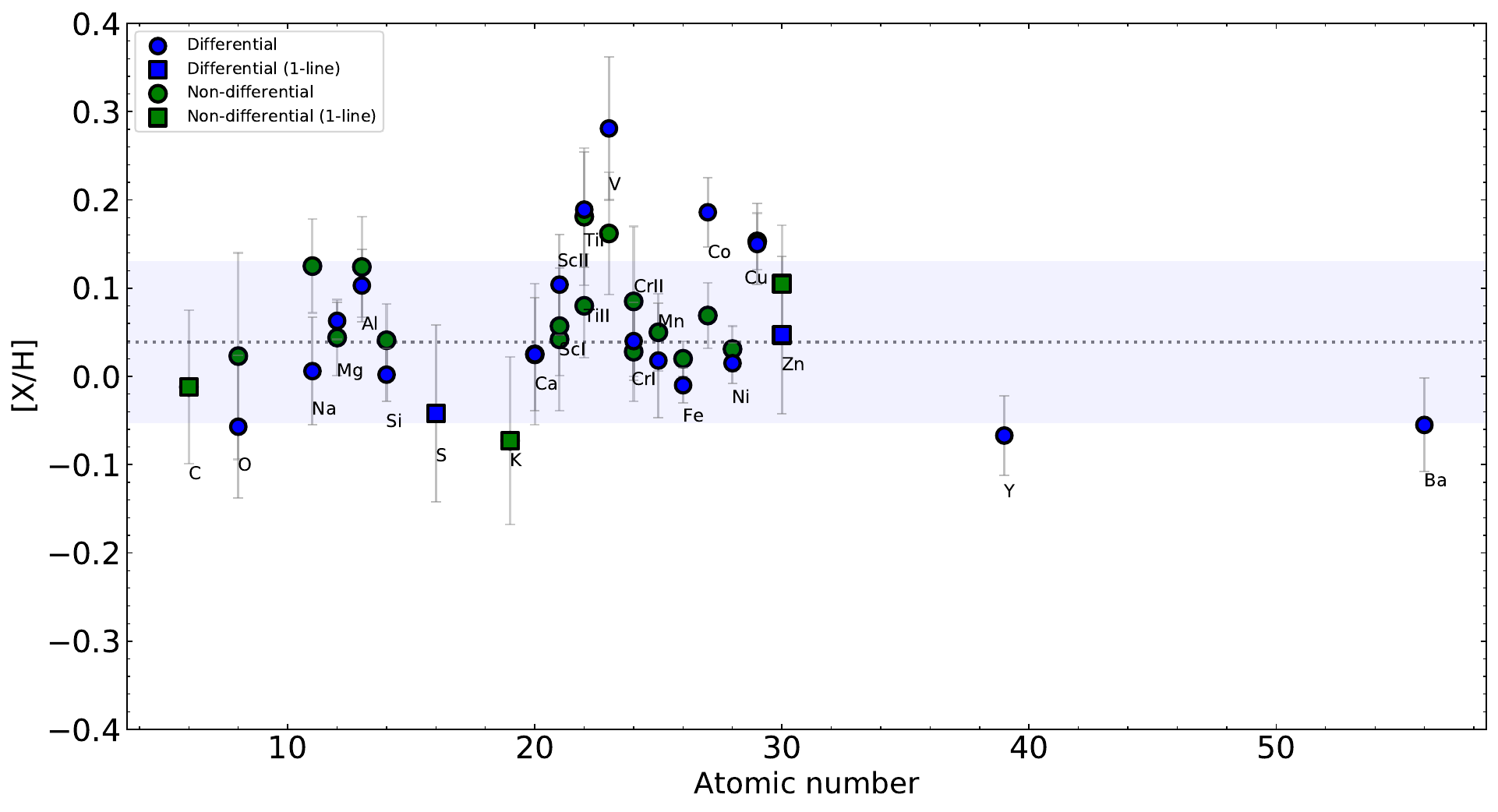}
\caption{Elemental abundances as a function of atomic number for TOI-3568 from the analysis of MAROON-X spectra. The methods are described in the legend, and each element's symbol is depicted in the plot. Dashed line marks the weighted average metallicity obtained from the differential abundances and the shaded area indicates the standard deviation.}
\label{fig:chemical-atomic-number}
\end{figure}

Table \ref{tab:abundances} lists both line-by-line differential abundances and non-differential values.
The uncertainties were determined considering the standard deviation of the mean abundances (for elements with three or more lines) as well as the contributions of the uncertainties on the atmospheric parameters. In order to calculate these contributions, we vary each atmospheric parameter in turn by $\pm$1$\sigma$ and calculate new abundances for all elements. We then subtract these new values from the original ones and obtain two differences caused by the variation of each atmospheric parameter. The contribution of each atmospheric parameter for the final uncertainty is taken as the maximum of these two differences. Finally, we add in quadrature the contributions of each of the four atmospheric parameters as well as the standard deviation of the mean abundances. For elements with only one or two lines, the latter is not considered. Figure \ref{fig:chemical-atomic-number} shows the abundances as a function of atomic number. There is a very good agreement between both sets of results, which are consistent within 1$\sigma$. We also notice the good agreement between the O abundances obtained from different indicators as well as between neutral and ionized species for Sc, Ti and Cr, which further shows that the ionization equilibrium achieved during the determination of the atmospheric parameters is robust. 

 \begin{table*}[h!]
  \small
      \caption{Strictly differential abundances (relative to the Sun) and non-differential abundances of TOI-3568}
\label{tab:abundances}
     \centering
         \begin{tabular}{l c c c c c c}
            \hline\hline

Element	&	[X/H] (dex) &	Error (dex)	&	N$_{lines}$	&	[X/H] (dex)	&	Error (dex) & N$_{lines}$	\\
	&	(line-by-line differential)	&	&		&	(non-differential)	&	& 	\\
\hline
Li	&	--	&	--	&	--	&	$\leq0.45$	&	--	&	1	\\
C   &   --  &   --  &   --  &   -0.012   &   0.087    &   1   \\
O I 7771-5 \AA	&	-0.057	&	0.081	&	2	&	0.023	&	0.117	&	3	\\
$[$O I$]$ 6300 \AA  &   --  &   --  &   --  &   0.015    &   0.063    &   1   \\
Na	&	0.006	&	0.061	&	4	&	0.125	&	0.053	&	3	\\
Mg	&	0.063	&	0.021	&	3	&	0.044	&	0.043	&	3	\\
Al	&	0.103	&	0.041	&	2	&	0.124	&	0.057	&	3	\\
Si	&	0.002	&	0.030	&	28	&	0.041	&	0.041	&	20	\\
S	&	-0.042	&	0.100	&	1	&	--	&	--	&	--	\\
K   &   --  &   --  &   --  &   -0.073  &   0.095   &   1   \\
Ca	&	0.025	&	0.064	&	16	&	0.025	&	0.080	&	8	\\
Sc I	&	--	&	--	&	--	&	0.042	&	0.081	&	2	\\
Sc II   &   0.104  &   0.057  &   8  &   0.057   &   0.056   &   8   \\
Ti I	&	0.189	&	0.065	&	31	&	0.181	&	0.078	&	19	\\
Ti II   &   --  &   --  &   --  &   0.080   &   0.059   &   6   \\
V	&	0.281	&	0.081	&	10	&	0.162	&	0.069	&	12	\\
Cr I	&	0.040	&	0.044	&	19	&	0.028	&	0.056	&	14	\\
Cr II   &   --  &   --  &   --  &   0.085   &   0.085   &  2   \\
Mn	&	0.018	&	0.065	&	7	&	0.050	&	0.044	&	6	\\
Fe 	&	-0.010	&	0.020	&	123	&	0.020	&	0.020	&	137	\\
Co	&	0.186	&	0.039	&	9	&	0.069	&	0.037	&	12	\\
Ni	&	0.015	&	0.023	&	46	&	0.031	&	0.026	&	35	\\
Cu	&	0.150	&	0.046	&	3	&	0.153	&	0.032	&	4	\\
Zn	&	0.047	&	0.089	&	1	&	0.105	&	0.066	&	1	\\
Y	&	-0.067	&	0.045	&	2	&	--	&	--	&	--	\\
Ba	&	-0.055	&	0.053	&	2	&	--	&	--	&	--	\\

    \hline
         \end{tabular} 
        \end{table*}  

\begin{figure}
\centering
\includegraphics[width=1.0\hsize]{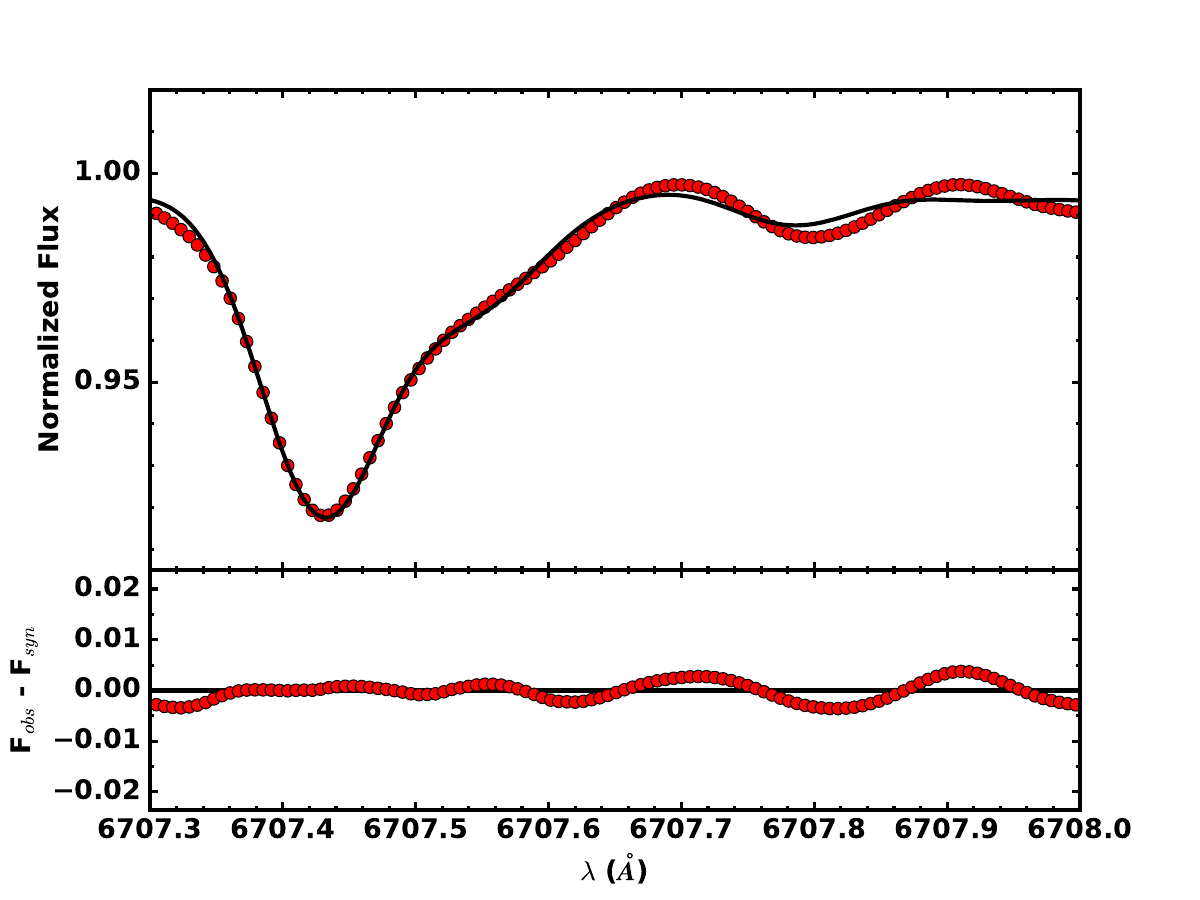}
\caption{Spectral synthesis for the Li~I feature at 6707.8~\AA~ for TOI-3568. \textit{Top panel:} Best fit between the observed (red circles) and synthetic (black line) spectra. The lithium line is not detected and we determined an upper limit of A(Li) $\leq$ 0.45. \textit{Bottom panel:} Residuals of the fit, which are within 0.5\%.}
\label{fig:lithium}
\end{figure}

We also determined the lithium abundance by performing a spectral synthesis of the Li~I feature at 6707.8~\AA. We adopted the line list from \cite{Ghezzi2010b} and a similar methodology, but with two differences: (1) we determined the Gaussian broadening (that considers the combined effects of the instrumental profile, stellar rotation and macroturbulence) using the Fe I line at 6703.567 \AA~ and (2) we kept the abundances of Fe, C and Si as free parameters (using the previously determined values as initial guesses) for the fit. As we can see in Figure \ref{fig:lithium}, the best fit shows no distinguishable Li feature and the residuals are smaller than 0.5\%. We are only able to determine an upper limit of A(Li) $\leq$ 0.45 and this value is consistent with those of stars with similar effective temperatures \citep[e.g.,][]{Ghezzi2010b, Ramírez2012}.

\subsection{Stellar population membership} 
\label{sec:stellarpopmembership}

\begin{figure}
\centering
\includegraphics[width=1.0\hsize]{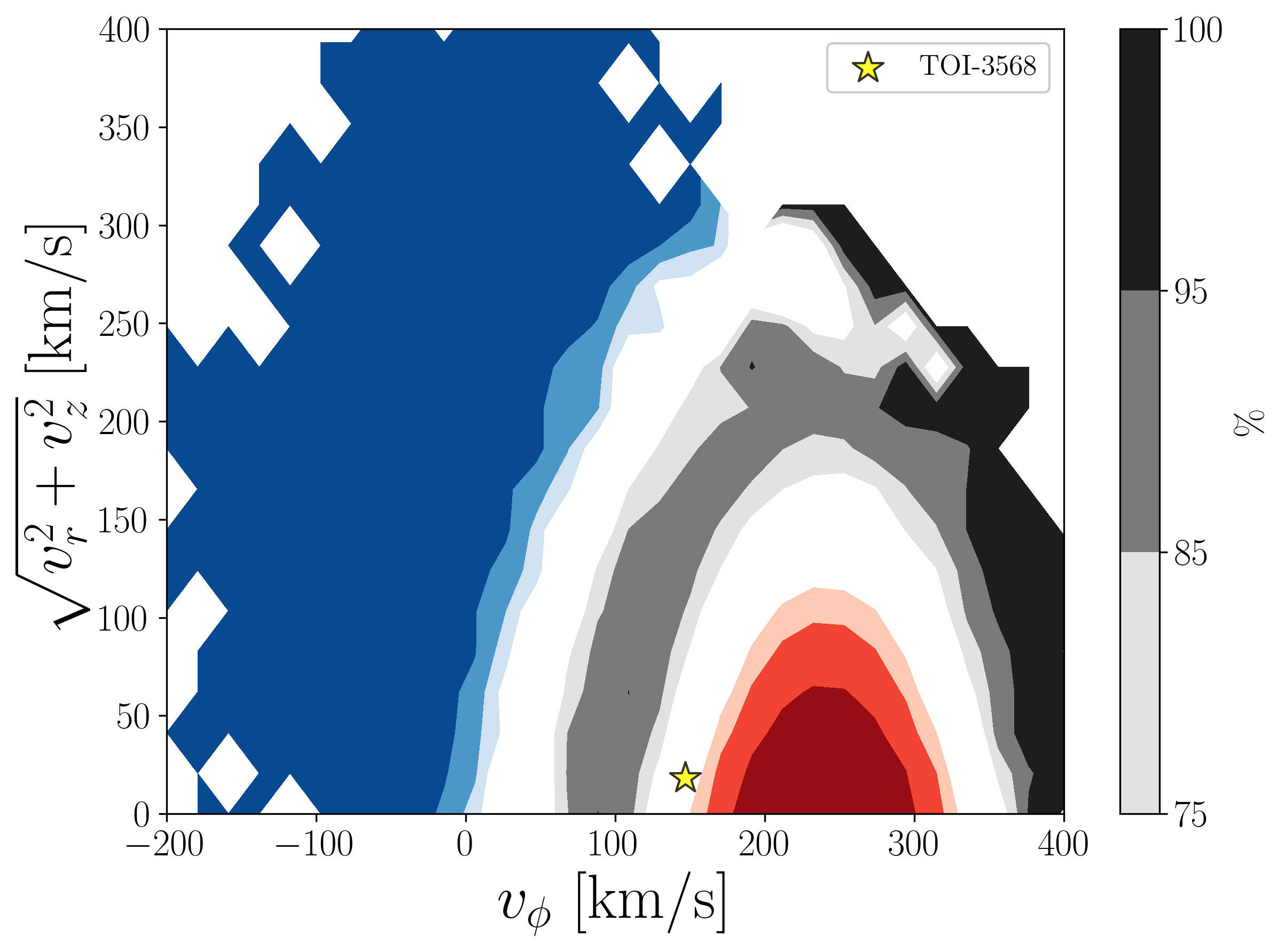}
\caption{Toomre diagram for the galactic components based on the \textsc{Galaxia} model. TOI-3568 is represented by the yellow star. The thin disk, thick disk, and halo are represented by the red, gray and blue colours, respectively. The varying darkness of shading indicates the fraction of each population at specific locations on the diagram (75\%, 85\% or 95\%).}
\label{fig:toomre}
\end{figure}

As we mention on the introduction, TOI-3568 was classified as a thick disk star by \cite{Carrillo2020}. To independently check this previous classification, we determined the galactic population membership in two ways. First, we performed a kinematic classification by computing the Toomre diagram (e.g., \citealt{bensby2003}). To identify regions dominated by each galactic population in the $v_{\phi}$ vs. $\sqrt{v_R^2 + v_z^2}$ plane, we utilized the \textsc{Galaxia} model \citep{sharma2011} with updated velocity distributions and local fractions of the thick disk and stellar halo (\citealt{amarante2020b}). This is illustrated in Figure \ref{fig:toomre}, where the thin disk, thick disk, and halo are represented by red, grey, and blue shaded areas, respectively. The kinematic characteristics of TOI-3568 are consistent with those typically associated with nearby thin-disk stars. We also applied a formalism similar to that of \citet{bensby2003} (see \citealt{Perottoni2021} for more details) to kinematically classify the galactic component of TOI-3568. Stars are classified as thick disk in this method when they show thick-disk-to-thin-disk (TD/D) membership ratios higher than 10, a condition not met by TOI-3568 ($TD/D = 1.9$).

Furthermore, we estimated the orbital properties of TOI-3568, following the description provided by \cite{Perottoni2021}. TOI-3568 is on prograde orbit with a high angular momentum $L_z = 1.2$ $\times 10
^{3}$~kpc\,km\,s$^{-1}$, mild eccentricity $e = 0.37$, and it reaches $Z_{\max} = 0.08$~kpc from the galactic plane. Additionally, TOI-3568 is situated in a region of $E$--$L_Z$ space predominantly occupied by disk stars. Although TOI-3568 exhibits an uncommon eccentricity, the other orbital parameters are similar to those of thin disk stars. Understanding the mechanism that led to this star's eccentricity is beyond the scope of this paper.

\begin{figure}
\centering
\includegraphics[width=1.0\hsize]{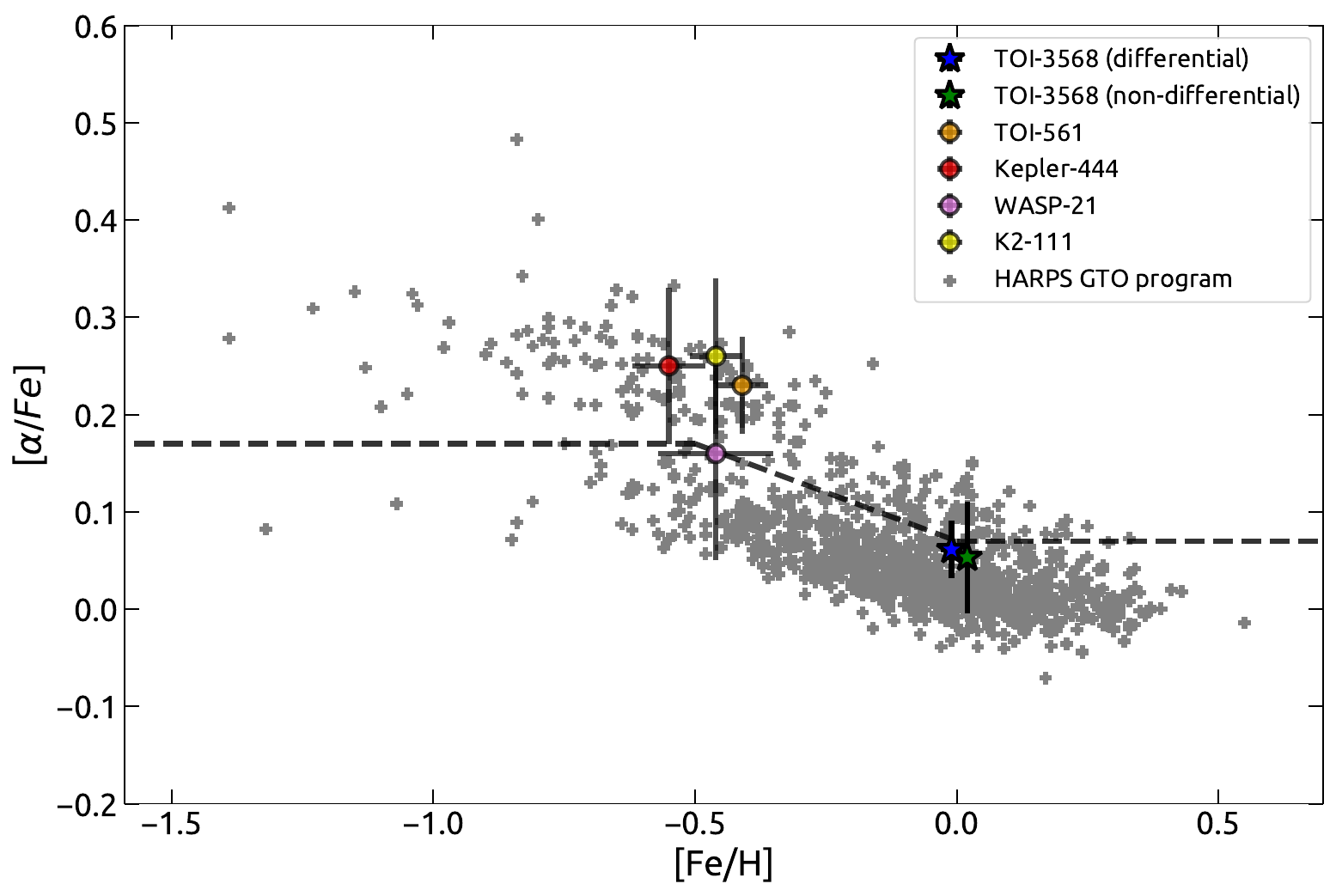}
\caption{Iron abundance vs. [$\alpha/{\rm Fe}$] for stars in the HARPS GTO program. The position of TOI-3568 is marked by a blue star (differential abundances) and by a green star (non-differential abundances). The black dashed line separates the galactic thin-disk (bottom) and thick-disk populations (top). For comparison, we also include the bona fide thick disk stars with transiting planets Kepler-444 \citep{campante2015}, TOI-561 \citep{Weiss2021}, K2-111 \citep{Mortier2020}, and WASP-21 \citep{Bouchy2010}.}
\label{fig:alpha-Fe}
\end{figure}

On the other hand, we also performed a classification of TOI-3568 based on its chemical composition. Generally, thick disk stars are metal-poor and enhanced in $\alpha$ elements \citep[e.g.,][]{Fuhrmann1998, reddy2006, Haywood2008, adibekyan2012}. Figure \ref{fig:alpha-Fe} shows the [$\alpha/{\rm Fe}$]\footnote{Here ``$\alpha$'' indicates the average abundance of Ca, Mg, Si, and Ti.} versus [Fe/H] for TOI-3568 in comparison with 1111 FGK dwarfs observed within the context of the HARPS GTO planet search program for which precise abundances of $\alpha$-elements are available \citep{adibekyan2012}. In this figure the objects are chemically separated, by the dashed line, between thick disk stars, overabundant in $\alpha$ elements and thin disk stars with less content of $\alpha$-elements for a given [Fe/H].

In line with the kinematic classification, the $\alpha$ content of TOI-3568, with a [$\alpha/{\rm Fe}$] value of $0.061\pm0.029$~dex from the differential method (and $[\alpha/{\rm Fe}]=0.05\pm0.06$~dex for the non-differential approach), does not exhibit significant enhancement for its metallicity. Therefore, TOI-3568 falls within the thin-disk region, albeit near the transition zone. For comparison, in Figure \ref{fig:alpha-Fe} we include bona fide thick disk stars with transiting planets for which there are alpha-element abundances available derived from high-resolution optical spectra. 

Additional support to our kinematic and chemical Galaxy population classification of TOI-3568 is based on its age. Generally, thick-disk field stars of the Milky Way are older than about 10–11 Gyr \citep[e.g.,][]{Bensby2005, reddy2006, Adibekyan2011}. Hence, the age of TOI-3568, estimated from the isochrones analysis (approximately in the range of 1-9 Gyr considering the errors in both methods, Sec. \ref{sec:mass_radius_age}) would be more consistent with those of thin-disk stars \citep[$\lesssim$ 9 Gyr, e.g.,][]{Bernkopf2001, Nissen2004, Kilic2017}. Moreover, as a further check to the ages from isochrones estimated above, we utilize the age-[Y/Mg] relation from \cite{TucciMaia2016} to obtain the age of TOI-3568 employing our measured abundances. This yields an age of $7.6\pm1.2$~Gyr, consistent with the values derived from the isochrones analysis. Therefore, our kinematic and chemical analysis indicate that TOI-3568 is a thin disk (or thin/thick disk transition object) rather than a thick disk star as indicated previously by \cite{Carrillo2020}.

\section{Planet detection and characterization}
\label{sec:planetcharacterization}

\subsection{Analysis of TESS photometry data}
\label{sec:analysisofTESSdata}

We analyzed the TESS data from sectors 15, 55, and 56 using methods outlined in \cite{Martioli2021}. This involved fitting a transit model along with a baseline polynomial within selected windows around each transit of TOI-3568~b. We selected six transits with a 30-minute cadence from sector 15 and thirteen transits with a 2-minute cadence from sectors 55 and 56. On average, each transit window contained 32 data points in sector 15 and 468 data points in sectors 55 and 56. This window size includes approximately twice as many out-of-transit data points on each side as in-transit ones. This was found to be a reasonable balance for a non-active star, providing enough baseline data to constrain transit parameters accurately and to allow for precise modeling of the baseline using a low-order polynomial. In this case, we use a first-order polynomial. As illustrated in Figure \ref{fig:toi3568_tess_lc}, we have also modeled the TESS photometry data with a Gaussian Process (GP) regression using a quasi-periodic kernel as in \cite{Martioli2022,Martioli2023}. The primary purpose of employing the GP in this analysis is for detrending, as it does not constrain any significant periodicity. Since the GP approach doesn't show significant improvements compared to the window approach, we opt for the latter for the rest of our analysis for the sake of simplicity.

Our transit model is calculated using the code \textsc{BATMAN} \citep{Kreidberg2015}, adopting a linear limb-darkening law. The quality of our ground-based WCWO photometry is not sufficient to constrain the limb-darkening on its own. To simplify, we adopted a single coefficient for both TESS and WCWO photometry, which is a reasonable approximation given the significant overlap between the TESS and r-band passes. The posterior distribution of transit parameters is sampled using a Bayesian Monte Carlo Markov Chain (MCMC) framework with the package \textsc{emcee} \citep{foreman2013}. We use uninformative priors for the transit parameters, shown in Table \ref{tab:planetsparams}.

\subsection{Joint analysis of RVs and photometry to obtain the system's parameters}
\label{sec:jointanalysis}

\begin{figure}
\centering
\includegraphics[width=1.0\hsize]{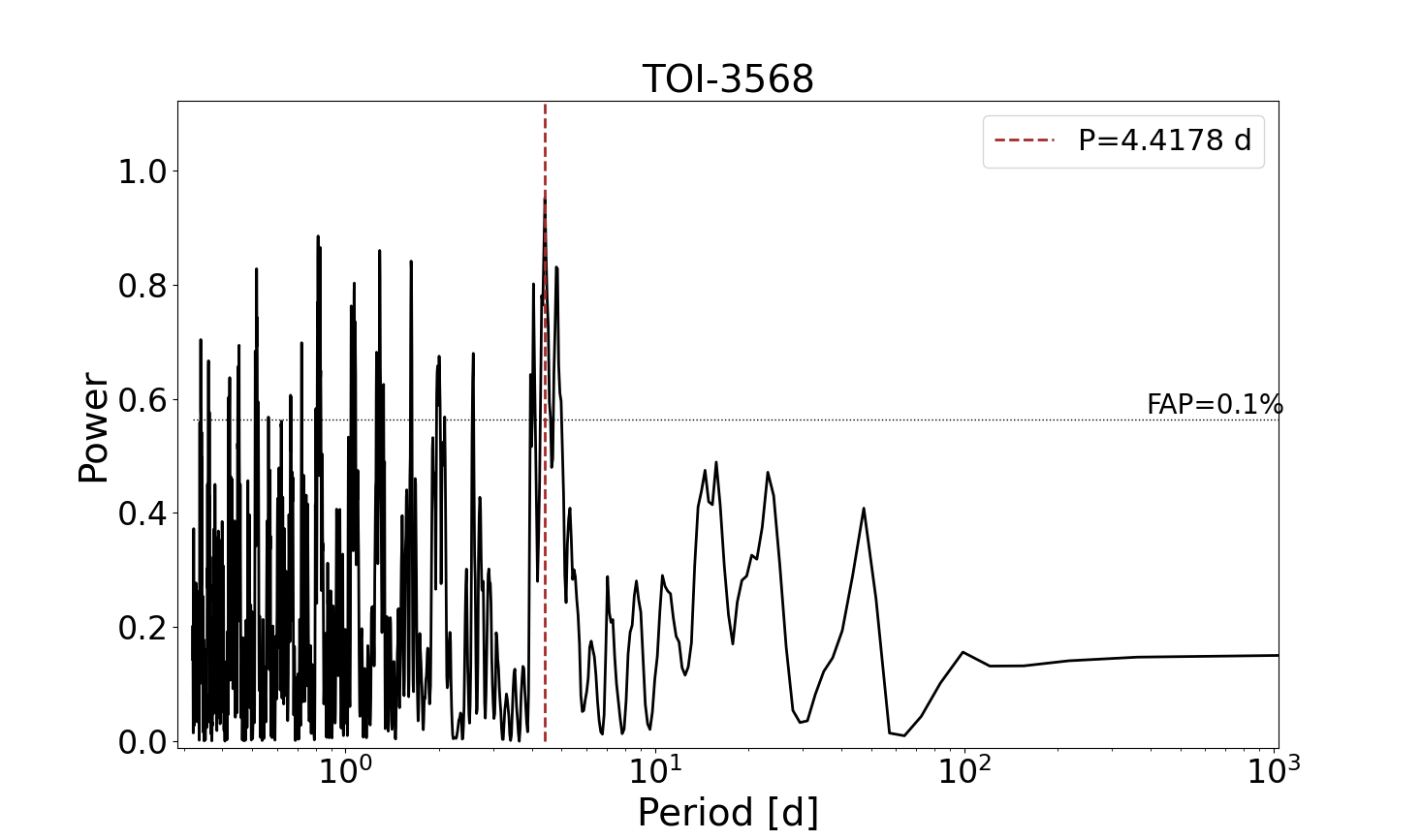}
\caption{Generalized Lomb-Scargle periodogram for TOI-3568 MAROON-X and SPIRou RVs data. The highest power is detected at 4.4178~d (dashed red vertical line) with ${\rm FAP}<10^{-14}$. }
\label{fig:TOI-3568_RV_GLS}
\end{figure}

We calculate the Generalized Lomb-Scargle \citep[GLS,][]{Zechmeister2009} periodogram for the MAROON-X and SPIRou RVs (see Tables \ref{tab:maroonx_rvs} and \ref{tab:spirou_rvs+blong}), as illustrated in Figure \ref{fig:TOI-3568_RV_GLS}. The highest power is detected at 4.4178~d with a false alarm probability (FAP) of ${\rm FAP}<10^{-14}$, coinciding precisely with the periodicity of transits observed in the TESS photometry data alone. This suggests that our RV data detects the signal of the star's reflex motion induced by the orbit of the planet TOI-3568~b.   

We therefore perform a Bayesian MCMC joint analysis of RVs and photometry data to determine the system's parameters employing the same approach as in \cite{Martioli2022,Martioli2023}. We calculate the log-likelihood for a model that includes both the transits and the RV orbit using a Keplerian model as in \cite{Martioli2010}, allowing for the simultaneous fitting of the TESS photometry data within the transit windows, the ground-based single-transit photometry data, and the RV data. The RV model includes an independent systemic velocity for each RV data set. In the case of MAROON-X blue and red data, this velocity represents only a systematic offset with respect to the template spectrum and therefore should be close to zero. On the other hand, the SPIRou systemic RV should be close to the real radial velocity of the system, although the value obtained in our analysis is not absolutely calibrated.
Initially, the parameters are fitted using an optimization least-squares (OLS) code with initial parameters obtained from an iterative preliminary analysis. We include a white noise jitter term for each RV dataset, which is fitted only in the OLS analysis. This is followed by sampling the posteriors using the Bayesian MCMC framework implemented with the \textsc{emcee} package. We run 20,000 iterations with 50 random walkers, discarding the first 5,000 samples as burn-in. The priors and posteriors for each parameter are presented in Table \ref{tab:planetsparams}, where the best-fit values are considered to be the mode of the distribution, and the errors are the 34th percentile on each side of the median.  Using these fit parameters and the stellar parameters from Table \ref{tab:stellarparams}, we derived other planetary quantities, which are also listed in Table \ref{tab:planetsparams}. Note that we employ bounded priors for the time of conjunction and orbital period, which have sufficiently wide bounds in the uniform distribution to ensure minimal impact on the posterior distribution. The best-fit orbit model for TOI-3568\,b shows an eccentricity of $0.035\pm0.021$, consistent within $2.5\sigma$ with a circular orbit, which is not uncommon for close-in exoplanets.

\begin{figure}
\centering
\includegraphics[width=1.0\hsize]{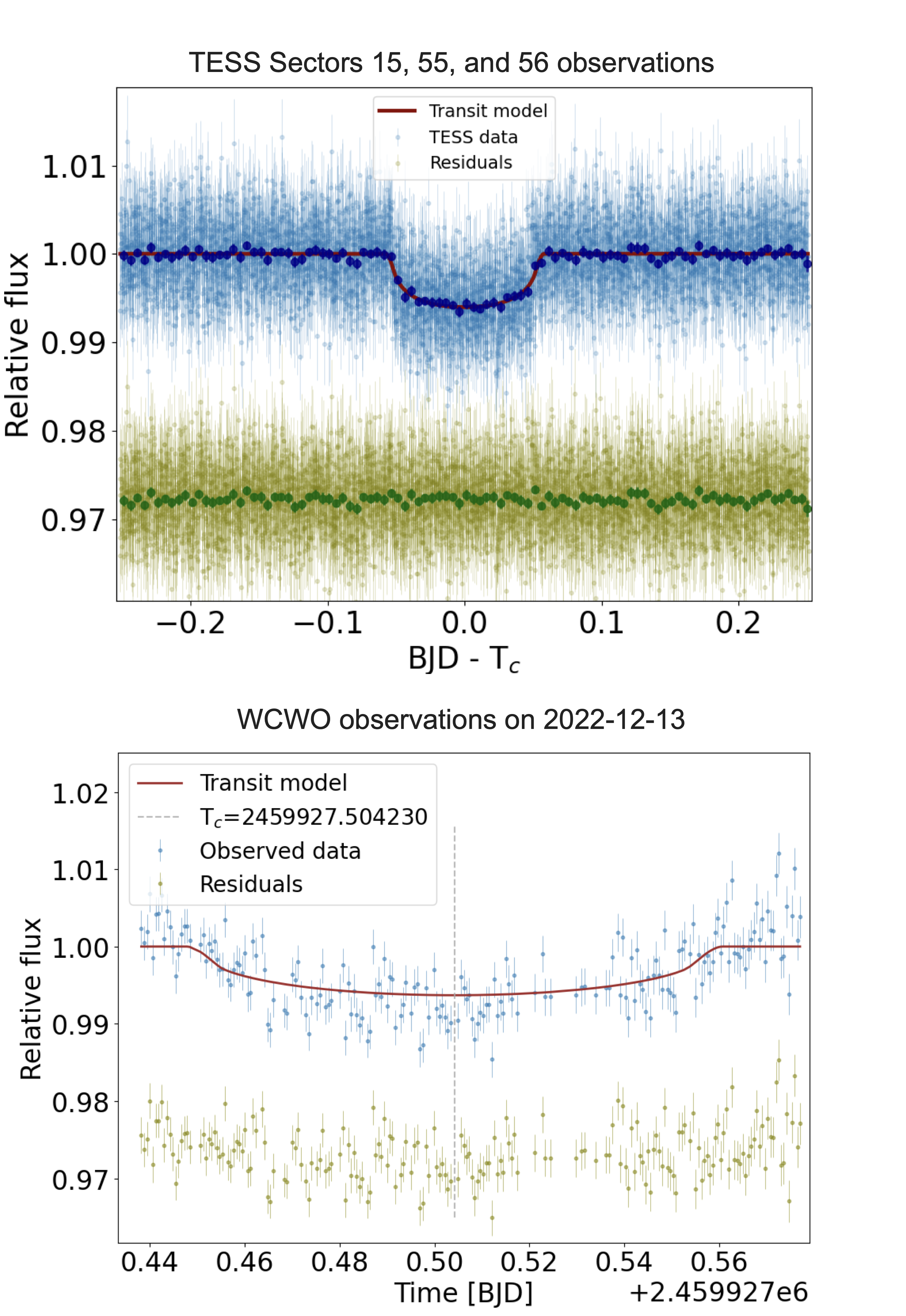}
\caption{Photometric observations of the transits of TOI-3568~b. The top panel shows the nineteen transits observed by TESS, while the bottom panel exhibits the single transit observed by our ground-based follow-up using the 0.7~m telescope at the Wellesley College Whitin Observatory.  The light blue points show the photometry data around the transits of TOI-3568~b, with the times in the top panel being relative to the central time of each transit. The red lines show the best-fit transit model and the green points show the residuals plus an arbitrary offset for better visualization. The dark blue and green points in the top panel represent the weighted averages of bins with a size of 0.01~d.}
\label{fig:TOI-3568_transitfit}
\end{figure}

\begin{figure*}
\centering
\includegraphics[width=0.95\hsize]{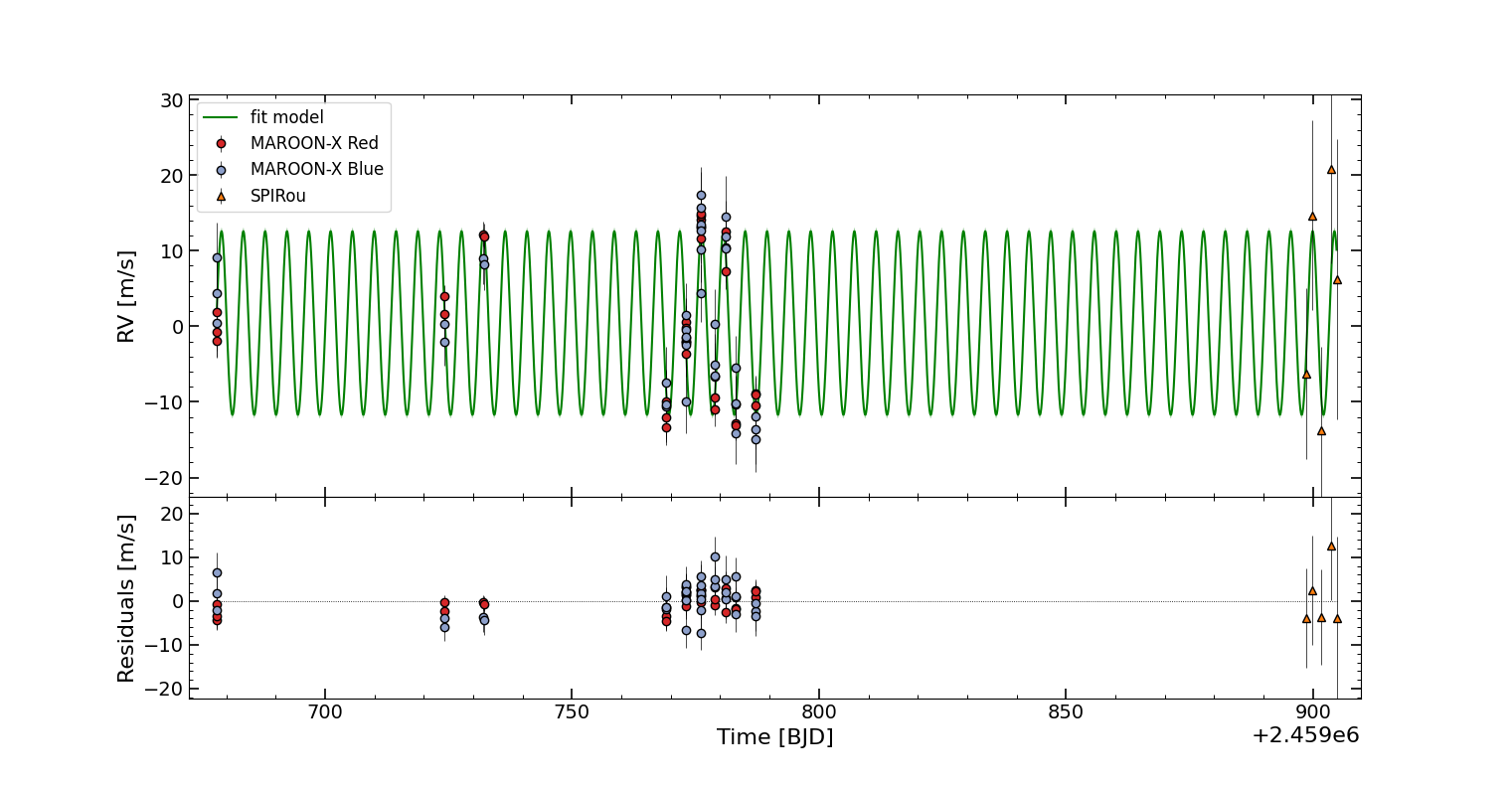}
\caption{TOI-3568 MAROON-X and SPIRou RVs. In the top panel, the red and blue points show the MAROON-X RV data for the red and blue channels, and the orange triangles show the SPIRou RV data. The green line shows the best-fit orbit model for TOI-3568~b. The bottom panel shows the residuals.}
\label{fig:TOI-3568_maroon-x+spirou_rvfit}
\end{figure*}

\begin{figure}
\centering
\includegraphics[width=0.95\hsize]{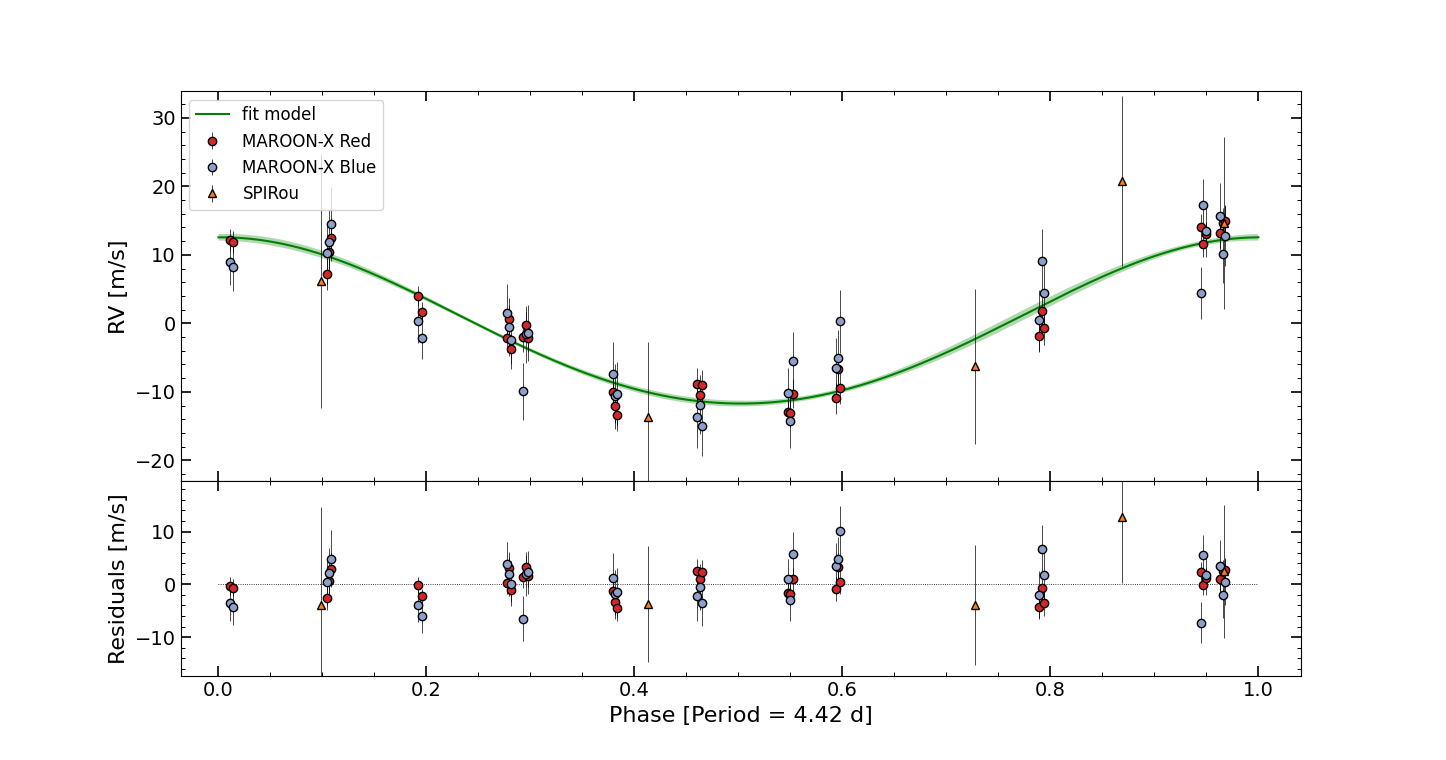}
\caption{TOI-3568 RVs in phase with the orbit of the planet. In the top panel, the red and blue points show the MAROON-X RV data for the red and blue channels, and the orange points show the SPIRou RV data. The green line shows the best-fit orbit model for TOI-3568~b. The bottom panel shows the residuals.}
\label{fig:TOI-3568_maroon-x+spirou_rvfit_phase}
\end{figure}

Figure \ref{fig:TOI-3568_transitfit} illustrates both the TESS and ground-based photometry data, and the best-fit model for the selected windows around the transits.  Figures \ref{fig:TOI-3568_maroon-x+spirou_rvfit} and \ref{fig:TOI-3568_maroon-x+spirou_rvfit_phase} illustrate the MAROON-X and SPIRou RV data and the orbit model. The final RMS of photometry residuals is 3.6~ppt, while for RV residuals, it is 3.5~\ms. The MCMC samples, their correlations, and the posterior distributions for each parameter are illustrated in Figure \ref{fig:TOI-3568_pairsplot} in  Appendix \ref{app:pairsplot}.

\begin{table*}
\centering
\caption{Priors and posteriors of the parameters of TOI-3568\,b obtained from a joint analysis of TESS and WCWO photometry, MAROON-X and SPIRou RVs.}
\label{tab:planetsparams}
\begin{tabular}{lcc}
\hline
Parameter & Prior & Posterior \\
\hline
time of conjunction, $T_{c}$ (BJD) & $\mathcal{U}(2459799,2459800)$ & $2459799.3834\pm0.0006$  \\
orbital period, $P$ (d) & $\mathcal{U}(4.415,4.420)$ & $4.417965\pm0.000005$  \\
eccentricity, $e$  & derived & $0.035\pm0.021$ \\
argument of periastron, $\omega$ (deg) &  derived & $-0.9\pm0.7$  \\
$e\sin{\omega}$ & $\mathcal{U}(-1,1)$ & $-0.03\pm0.04$ \\
$e\cos{\omega}$ & $\mathcal{U}(-1,1)$ & $0.02\pm0.04$ \\
normalized semimajor axis, $a/R_{\star}$ & $\mathcal{U}(1,20)$& $13.1^{+0.8}_{-1.4}$ \\
\tablefootmark{a} semimajor axis, $a_{p}$ (au) & derived & $0.0485\pm0.0004$ \\
orbital inclination, $i_{p}$  (deg) &  $\mathcal{U}(85,90)$ & $89.2^{+1.1}_{-1.3}$  \\
transit duration, $t_{\rm dur}$  (h) & derived &  $2.45\pm0.14$ \\
impact parameter, $b$  & $\mathcal{U}(0,1)$ &  $<0.50$ \\
planet-to-star radius ratio, $R_{p}/R_{\star}$  & $\mathcal{U}(0,1)$ & $0.067\pm0.003$\\
planet radius, $R_{p}$ (\RJ) & derived &  $0.483\pm0.024$ \\
planet radius, $R_{p}$ (\RN)  & derived &  $1.37\pm0.07$ \\
planet radius, $R_{p}$ (\RE) & derived  &  $5.30\pm0.27$  \\
velocity semi-amplitude, $K_{p}$ (m\,s$^{-1}$) & $\mathcal{U}(0,1000)$ &  $12.1\pm0.4$ \\
planet mass, $M_{p}$ (\MJ) & derived & $0.083\pm0.003$ \\
planet mass, $M_{p}$ (\MN) & derived & $1.54\pm0.06$ \\
planet mass, $M_{p}$ (\ME) & derived & $26.4\pm1.0$ \\
planet bulk density, $\rho_{p}$ (g\,cm$^{-3}$) & derived & $0.98\pm0.15$ \\
 \tablefootmark{b} equilibrium temperature, $T_{\rm eq}$ (K) & derived & $899\pm12$\\
linear limb dark. coef., $u_{0}$  & $\mathcal{U}(0,1)$ & $0.85\pm0.11$ \\
\tablefootmark{c} Systemic RV (MAROON-X, Blue), $\gamma_{\rm MAROONX\,Red}$ (m\,s$^{-1}$) & $\mathcal{U}(-\infty,+\infty)$ & $-1.2\pm0.7$  \\
\tablefootmark{c} Systemic RV (MAROON-X, Red), $\gamma_{\rm MAROONX\,Blue}$ (m\,s$^{-1}$) & $\mathcal{U}(-\infty,+\infty)$ & $-1.4\pm0.4$  \\
Systemic RV (SPIRou, NIR), $\gamma_{\rm SPIRou}$ (m\,s$^{-1}$) & $\mathcal{U}(-\infty,+\infty)$ & $-97667\pm6$  \\
MAROON-X Blue RV jitter (m\,s$^{-1}$) & \multicolumn{2}{c}{2.6} \\ 
MAROON-X Red RV jitter (m\,s$^{-1}$) & \multicolumn{2}{c}{1.0} \\ 
SPIRou RV jitter (m\,s$^{-1}$) & \multicolumn{2}{c}{0.0} \\ 
RMS of MAROON-X Blue RV residuals (m\,s$^{-1}$) & \multicolumn{2}{c}{2.2} \\
RMS of MAROON-X Red RV residuals (m\,s$^{-1}$) & \multicolumn{2}{c}{3.9} \\
RMS of SPIRou RV residuals (m\,s$^{-1}$) & \multicolumn{2}{c}{6.5}  \\
RMS of RV residuals (m\,s$^{-1}$) & \multicolumn{2}{c}{3.5} \\
RMS of TESS flux residuals (ppt) & \multicolumn{2}{c}{3.6} \\ 
RMS of WCWO flux residuals (ppt) & \multicolumn{2}{c}{3.5} \\ 
\hline
\end{tabular}
\tablefoot{
\tablefoottext{a}{semi-major axis derived from the fit period and Kepler's law.}
\tablefoottext{b}{assuming a uniform heat redistribution and an arbitrary geometric albedo of 0.1.}
\tablefoottext{c}{the MAROON-X RVs are relative to the template and, therefore, the systemic RVs represent a systematic offset relative to our orbital fit.}
}
\end{table*}

\subsection{Limits on additional planets from TESS photometry}
\label{sec:injrec}

We explored the detection limits of additional transiting planets by performing an injection-recovery test in the TESS light curve of TOI-3568. To do so, we employed the 2-min TESS PLD photometry residuals from sectors 55 and 56 obtained after removing the baseline GP model multiplied by the best-fit transit model for TOI-3568 b determined in Section \ref{sec:analysisofTESSdata}.

\begin{figure}
\centering
\includegraphics[width=1.0\hsize]{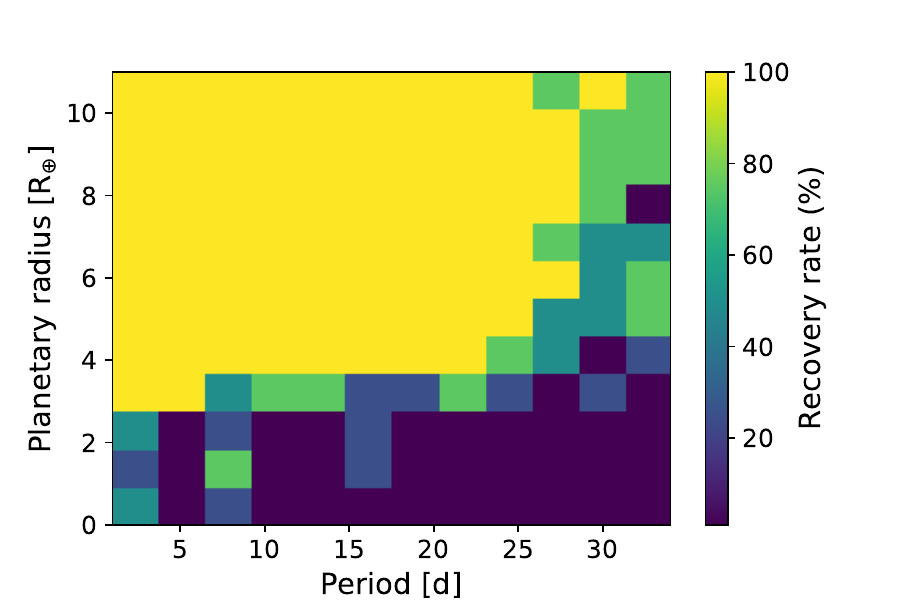}
\caption{Injection-recovery test performed on the TESS PLD light curve residuals from sectors 55 and 56 of TOI-3568 to check the detectability of additional planets in the system. High and low recovery rates are presented in light yellow-green and dark blue colors, respectively. From this detectability map, we can conclude that the existence of additional planets with radii $\gtrsim$4.0~\RE\ and periods $\lesssim$25~d should be excluded, given that the recovery rates range from 80 to 100\%. However, small planets with sizes $\lesssim$4.0~\RE\ would remain undetected for almost the entire set of periods investigated.}
\label{fig:injrecfig}
\end{figure}

We used the BATMAN code \citep{Kreidberg2015} to generate synthetic transit signals that were injected into the TESS PLD photometry residuals. For all of these simulated planets, a limb-darkening linear law, equatorial transits ($b = 0$) and circular orbits ($e = 0$) were assumed. The value adopted for the limb-darkening coefficient, $u_{0} = 0.85$, was extracted from Table \ref{tab:planetsparams}. We surveyed the planetary radius–orbital period parameter space, $R_P$--$P$, in the ranges of 0--11~\RE\  with steps of 1~\RE\  and 1--34~d with a 3~d step, respectively, adopting a multi-phase approach that allows five different values of $T_{c}$ for each $R_P$--$P$ combination. As in previous works \citep[see][]{Jofre2021}, to detect the injected signals, we ran the Transit Least Squares code \citep[TLS;][]{hippke2019b}, an optimized algorithm to search for periodic transits from time-series photometry. A positive planet detection was considered when the recovered orbital period is within 5$\%$ of any half multiple of the injected period. In Figure \ref{fig:injrecfig}, we present the detectability map resulting from our injection-recovery test.

From this plot it can be seen that planets with a radius larger than $\sim$4.0~\RE\ and periods $\lesssim$25~d have recovery rates of more than 80$\%$, hence, we can exclude the presence of such additional objects in the system. For similar-sized planets but orbital periods longer than 25~d, the chances of detection are between 0 and 100$\%$. As we consider longer periods, fewer transits of a given planet are expected. Here, the chances of detection can drop abruptly, from high to low percentages, if one or more transits are not detected (for example, if the event falls in the gap between orbits). In the space parameter corresponding to planet size smaller than $\sim$4.0~\RE, most of the recovery rates are lower than 20$\%$. This indicates that still might exist small planetary objects that would remain undetected in the present data.

\section{Discussions}
\label{sec:discussions}

\subsection{Characterization of TOI-3568~b}
\label{sec:characterizationtoi3568b}

\begin{figure}
  \sidecaption
   \includegraphics[width=1.0\hsize]{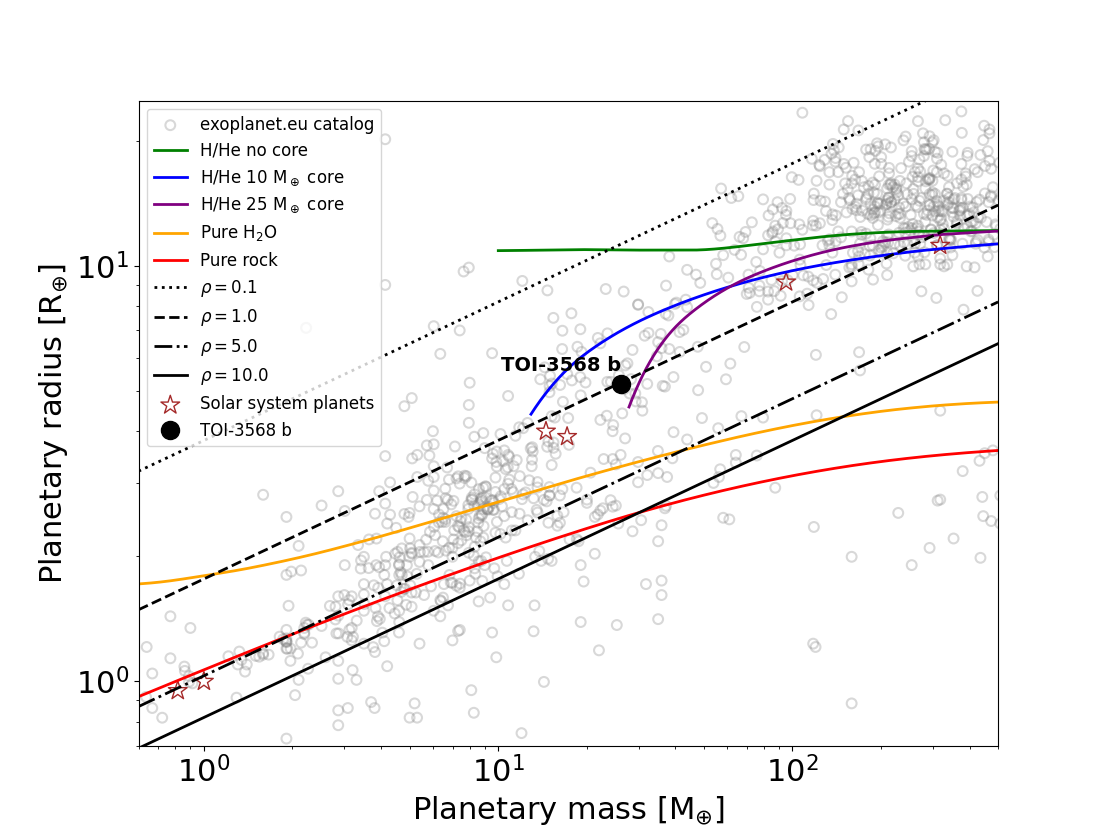}
      \caption{ Mass-radius diagram. 
      The grey circles show the exoplanet data from \texttt{exoplanets.eu} and the red stars show the solar system planets. The black point corresponds to the measured value for TOI-3568~b. Internal structure models for planets with pure rock and pure ice (H$2$O) compositions from \cite{Zeng2019} are depicted by the solid red and orange lines, respectively. Additionally, evolutionary models for H/He dominated planets from \cite{Fortney2007} are shown for a pure H/He without a core (green line), and for a core of 10~\ME\ (blue line) and 25~\ME\ (purple line) of heavier elements. Iso-density curves are also represented by the black lines.
      }
    \label{fig:massradiusdiagram}
  \end{figure}

Considering the planet-to-star radius ratio of $R_{p}/R_{\star}=0.067\pm0.003$ and the stellar radius obtained from our stellar analysis, we derive the true physical radius of TOI-3568\,b as $5.30\pm0.27$~\RE, which is approximately $1.37$ times the size of Neptune. The RV semi-amplitude of $12.1\pm0.4$~\ms\ implies a planet mass of $26.4\pm1.0$~\ME, approximately 50\% larger than Neptune's mass. We estimate a bulk density of $0.98\pm0.15$~g\,cm$^{-3}$.  In Figure \ref{fig:massradiusdiagram} we show the mass-radius diagram for the known exoplanets with masses in the range 0.5-500~\ME\ and radii in the range 0.7-25~\RE.  Comparison of the evolutionary models by \cite{Fortney2007} for Hydrogen-Helium (H/He) rich planets at 0.045~au and ages between 1 and 10~Gyr reveals that TOI-3568\,b is likely a H/He-dominated planet with a core of heavier elements, with a mass between 10 and 25~\ME.

With an orbital period of 4.4~days and at an orbital distance of $0.0485\pm0.0004$~au, we estimated the equilibrium temperature for TOI-3568\,b as in \cite{heng2013}, assuming a uniform heat redistribution and an arbitrary geometric albedo of 0.1, which gives $T_{\rm eq}=899\pm12$~K. Therefore this planet belongs to the hot super-Neptune class, a rare type of planet. 

\subsection{TOI-3568\,b in the sub-Jovian desert}
\label{sec:desert}

\begin{figure}
\includegraphics[width=1.0\hsize]{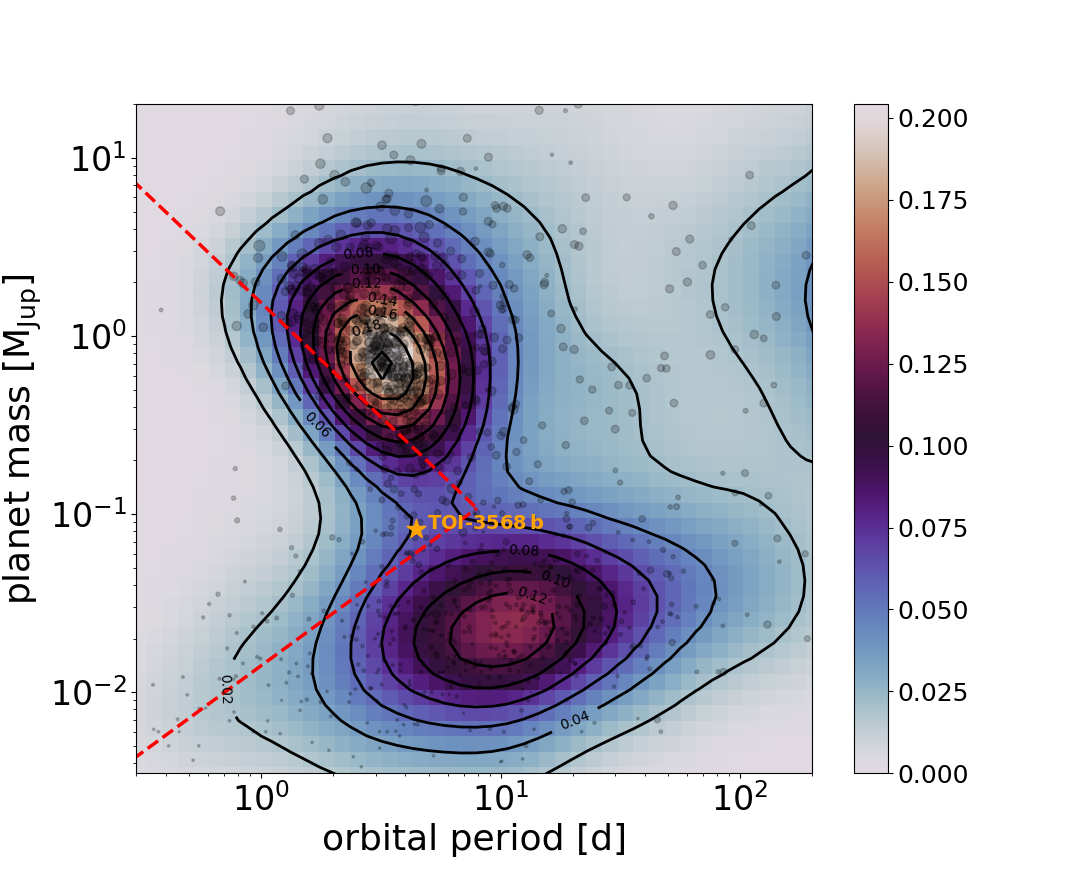}
\caption{ Mass-period diagram. The grey circles show the exoplanet data from \texttt{exoplanet.eu}, where the size of the circles is proportional to the planetary radius. The color map depicts the planet detection rate in percentage normalized to the sum of all known exoplanets in the mass range between 0.003~\MJ\ and 20~\MJ\, and orbital periods between 0.3~d and 200~d. The red dashed lines show the innermost limits of the sub-Jovian desert as defined by \cite{Mazeh2016}. The orange star shows TOI-3568~b, which lies within the desert right in the transition between the two populations of the most frequent types of exoplanets, hot-Jupiters (top) and warm super-Earths (bottom).}
\label{fig:massperioddiagram}
\end{figure}

In Figure \ref{fig:massperioddiagram}, we explore the mass-period diagram, illustrating the population of exoplanets. The color contrasts in this plot represent the detection rate rather than the actual occurrence rate, which ideally should reflect the fundamental physics of planet formation. However, selection effects, primarily due to the limited sensitivity of radial velocity and transit surveys, are significant only in the regime of long periods and small masses. Therefore, the scarcity of planets at short periods and masses above a few hundredths of a Jupiter mass should reflect the intrinsic planetary population in this region. We highlight the boundaries of the sub-Jovian desert as defined by \cite{Mazeh2016}. TOI-3568\,b lies within the boundaries of the desert, albeit in a region that has a low detection rate, yet is not entirely devoid of planets. TOI-3568~b falls near the lower boundary of the sub-Jovian desert, which is thought to be caused by photoevaporation \citep{Owen2018}. As pointed out by \cite{Frame2023}, this lower boundary is becoming more blurred as we detect more planets around that region, raising the question whether this need to be reevaluated. Regardless of the existence of a real desert in this region, TOI-3568\,b is notably positioned right within the transition between the populations of hot-Jupiters and super-Earths, thus having potential importance in investigating the origin of this natural segmentation of planet populations.

\subsection{Evaporation status of TOI-3568~b}
\label{sec:photoevaporation}

Photoevaporation is thought to play an important role in the distribution of planets in the mass-period diagram \citep{Owen2018,Mordasini2020,Vissapragada2022}. However, this diagram does not take into account different stellar types. To assess the evaporation status of TOI-3568~b, we plot it on the energy diagram proposed by \cite{lecavalier2007}. This diagram explores the relationship between the potential energy of the planet and the extreme ultraviolet (EUV) luminosity it receives from its parent star. These two competing sources of energy control the evaporation rate of the planet throughout its lifetime.

As in Eq. 10 of \cite{lecavalier2007}, we consider the planet's potential energy E$_{\rm p'}$, including tidal forces from the gravitational interaction with the parent star, and calculate the EUV luminosity following his recipe.  For simplicity, we assume a constant mean EUV luminosity and adopt a correction factor of $\gamma=6$ to account for the time variation in the energy flux received by the planet. This approximation becomes more problematic for very young systems and very hot stars, but both types of objects account for a small fraction of the exoplanets that we present in our analysis.

\begin{figure}
\includegraphics[width=1.0\hsize]{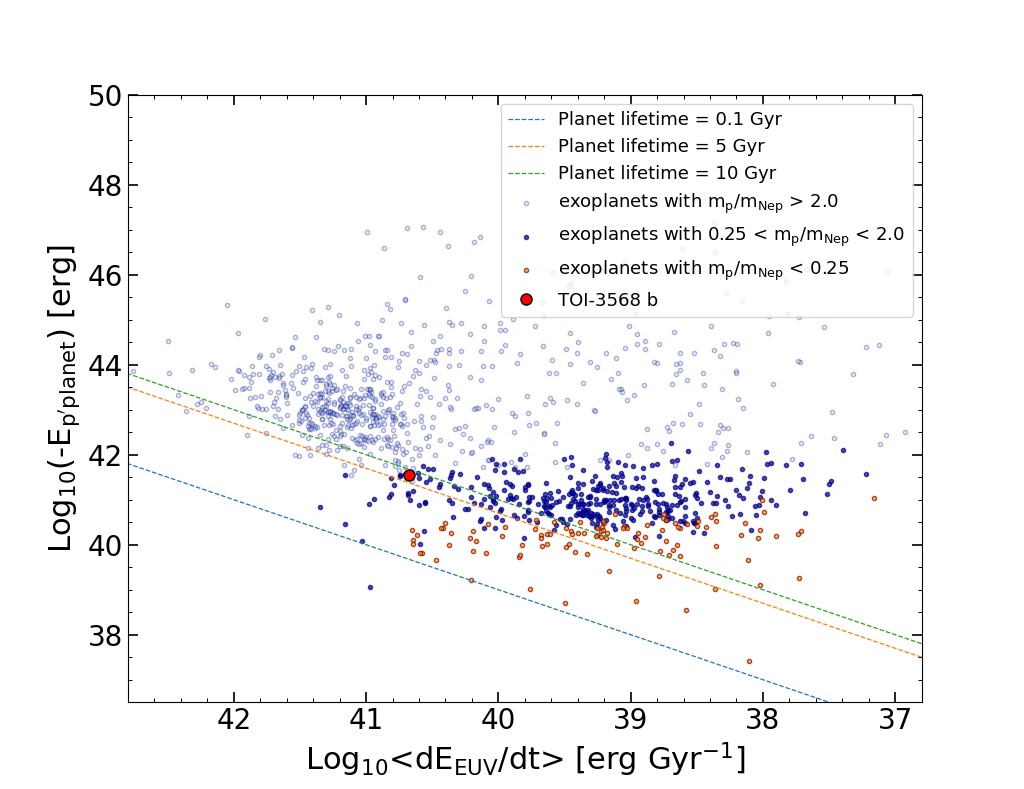}
\caption{ Diagram of planet potential energy versus mean EUV luminosity for transiting exoplanets from the \texttt{exoplanet.eu} catalog with measured masses. Exoplanets are represented by different symbols for three mass ranges: ${\rm M}_p > 2{\rm M}_{\rm Nep}$ (blue hollow circles), $0.25{\rm M}_{\rm Nep} < {\rm M}_p < 2{\rm M}_{\rm Nep}$ (blue filled circles), and ${\rm M}_p < 0.25{\rm M}_{\rm Nep}$ (orange circles).  TOI-3568~b is highlighted by the red circle. Dashed lines represent planet lifetimes of 0.1 Gyr (blue), 5 Gyr (orange), and 10 Gyr (green).}
\label{fig:energydiagram}
\end{figure}

Figure \ref{fig:energydiagram} reproduces the energy plot from \cite{lecavalier2007}, showing up-to-date exoplanet data from the \url{exoplanet.eu} catalog for transiting planets with measured masses and highlighting planets in three mass regimes: Jupiter-like (${\rm M}_p > 2{\rm M}_{\rm Nep}$), Neptune-like ($0.25{\rm M}_{\rm Nep} < {\rm M}_p < 2{\rm M}_{\rm Nep}$), and Earth-like (${\rm M}_p < 0.25{\rm M}_{\rm Nep}$). TOI-3568~b stands out as one of the super-Neptunes (${\rm M}_p = 1.54\pm0.06$~\MN) with the highest levels of EUV luminosity, receiving about  $dE_{\rm EUV}/dt > 10^{40.6}$~erg\,Gyr$^{-1}$. The dashed lines in Figure \ref{fig:energydiagram} represent planet lifetimes of 0.1, 5, and 10 Gyr. The regions below these lines are considered evaporation-forbidden regions, where planets receive more EUV energy than is needed to fill their potential well, and thus would evaporate in less than 0.1, 5, or 10 Gyr, respectively.

As in the mass-period diagram, this energy diagram also reveals two distinct populations of exoplanets: one population consists of more massive planets with high potential energy and high EUV luminosity. These planets accumulate in the top-left part of the diagram, above the forbidden region at 5 Gyr, with the lower boundary clearly sculpted by photoevaporation. A second population of less massive planets (M${p}<2$M$_{\rm Nep}$) is accumulated in a lower EUV luminosity regime in the bottom-right part of the diagram. TOI-3568~b lies at the transition between these two populations, making it difficult to clearly classify it into one category or the other. Although being in a region that is not significantly populated, TOI-3568~b is above a lifetime of 5~Gyr, which is consistent with the mature age of the system. However, the question remains: why is this region less populated? As pointed out by \cite{Owen2018}, there are likely other formation mechanisms sculpting the lower mass end, causing this natural gap in the population. TOI-3568~b appears to be an important planet for probing this gap. However, a more detailed analysis of this topic is beyond the scope of this paper.

\section{Conclusions}
\label{sec:conclusions}

We report the discovery of the transiting exoplanet TOI-3568~b, an inflated hot super-Neptune situated in the sub-Jovian desert. Using the TESS and ground-based photometry, MAROON-X optical spectra, and SPIRou NIR spectropolarimetry, we determined the orbit of the planet and the physical parameters of the system. The star is a quiet and mature K dwarf with an effective temperature of $4969\pm45$~K, with nearly solar metallicity. Our analysis identifies this star as part of the transitional population between the galactic thin and thick disk, exhibiting characteristics more consistent with the thin disk population.  Although our observations did not confirm this candidate as a member of the galactic thick disk, it emerged as an interesting discovery of a rare super-Neptune situated in a region of the mass-period diagram with a low occurrence rate of planets. TOI-3568\,b is likely an H/He-dominated planet with a core of heavier elements with a mass between 10 and 25~\ME, and it lies at the lowest point between the two significant populations of hot-Jupiters and super-Earths.  We analyzed the photoevaporation status of TOI-3568~b, finding that this planet experiences a high regime of EUV luminosity for its mass range, making it one of the planets with the highest EUV luminosities among those with a mass M$_{p}<2$~\MN. However, this planet is not in an evaporation-forbidden region, as its status is still consistent with a planet having an evaporation lifetime exceeding 5 Gyr. Perhaps the most interesting aspect of this planet is that it is not a common type of planet. It lies in a transition region in both the mass-period and energy diagrams, an area with a dearth of planets that cannot be explained by photoevaporation. 

\begin{acknowledgements}

This work is based on observations obtained at the Gemini Observatory, which is operated by the Association of Universities for Research in Astronomy, Inc., under a cooperative agreement with the NSF on behalf of the Gemini partnership:  the US National Science Foundation (NSF), the Canadian National Research Council (NRC), the Chilean Agencia Nacional de Investigación y Desarrollo (ANID), the Brazilian Ministério da Ciência, Tecnologia e Inovação, the Argentinean Ministerio de Ciencia, Tecnología e Innovación, and the Korea Astronomy and Space Institute (KASI). 

This work is based on observations obtained at the Canada-France-Hawaii Telescope (CFHT) which is operated from the summit of Maunakea by the National Research Council of Canada, the Institut National des Sciences de l'Univers of the Centre National de la Recherche Scientifique of France, and the University of Hawaii. Based on observations obtained with SPIRou, an international project led by Institut de Recherche en Astrophysique et Plan\'etologie, Toulouse, France.

E.M. acknowledges funding from Funda\c{c}\~{a}o de Amparo \`{a} Pesquisa do Estado de Minas Gerais (FAPEMIG) under project number APQ-02493-22 and research productivity grant (PQ) number 309829/2022-4 awarded by the National Council for Scientific and Technological Development (CNPq), Brazil.

The University of Chicago group acknowledges funding for the MAROON-X project from the David and Lucile Packard Foundation, the Heising-Simons Foundation, the Gordon and Betty Moore Foundation, the Gemini Observatory, the NSF (award number 2108465), and NASA (grant number 80NSSC22K0117). The Gemini observations are associated with programs ID GN-2022A-Q-207/-Q-113.

Funding for the TESS mission is provided by NASA’s Science Mission Directorate. We acknowledge the use of public TESS data from pipelines at the TESS Science Office and at the TESS Science Processing Operations Center. Resources supporting this work were provided by the NASA High-End Computing (HEC) Program through the NASA Advanced Supercomputing (NAS) Division at Ames Research Center for the production of the SPOC data products. DR was supported by NASA under award number NNA16BD14C for NASA Academic Mission Services. TESS data presented in this paper were obtained from the Mikulski Archive for Space Telescopes (MAST) at the Space Telescope Science Institute.

This work has made use of data from the European Space Agency (ESA) mission Gaia (\url{https://www.cosmos.esa.int/gaia}), processed by the Gaia Data Processing and Analysis Consortium (DPAC, \url{https://www.cosmos.esa.int/web/gaia/dpac/consortium}). Funding for the DPAC has been provided by national institutions, in particular, the institutions participating in the Gaia Multilateral Agreement.

This work made use of \textsc{tpfplotter} by J. Lillo-Box (publicly available in www.github.com/jlillo/tpfplotter), which also made use of the python packages \textsc{astropy}, \textsc{lightkurve}, \textsc{matplotlib} and \textsc{numpy}.

KKM acknowledges support from the New York Community Trust Fund for Astrophysical Research.

\end{acknowledgements}

%
%

\bibliographystyle{aa}

\bibliography{bibliography}

\begin{appendix}
\section{Posterior distributions of model parameters}
\label{app:pairsplot}

\begin{figure*}
\centering
\includegraphics[width=1.0\hsize]{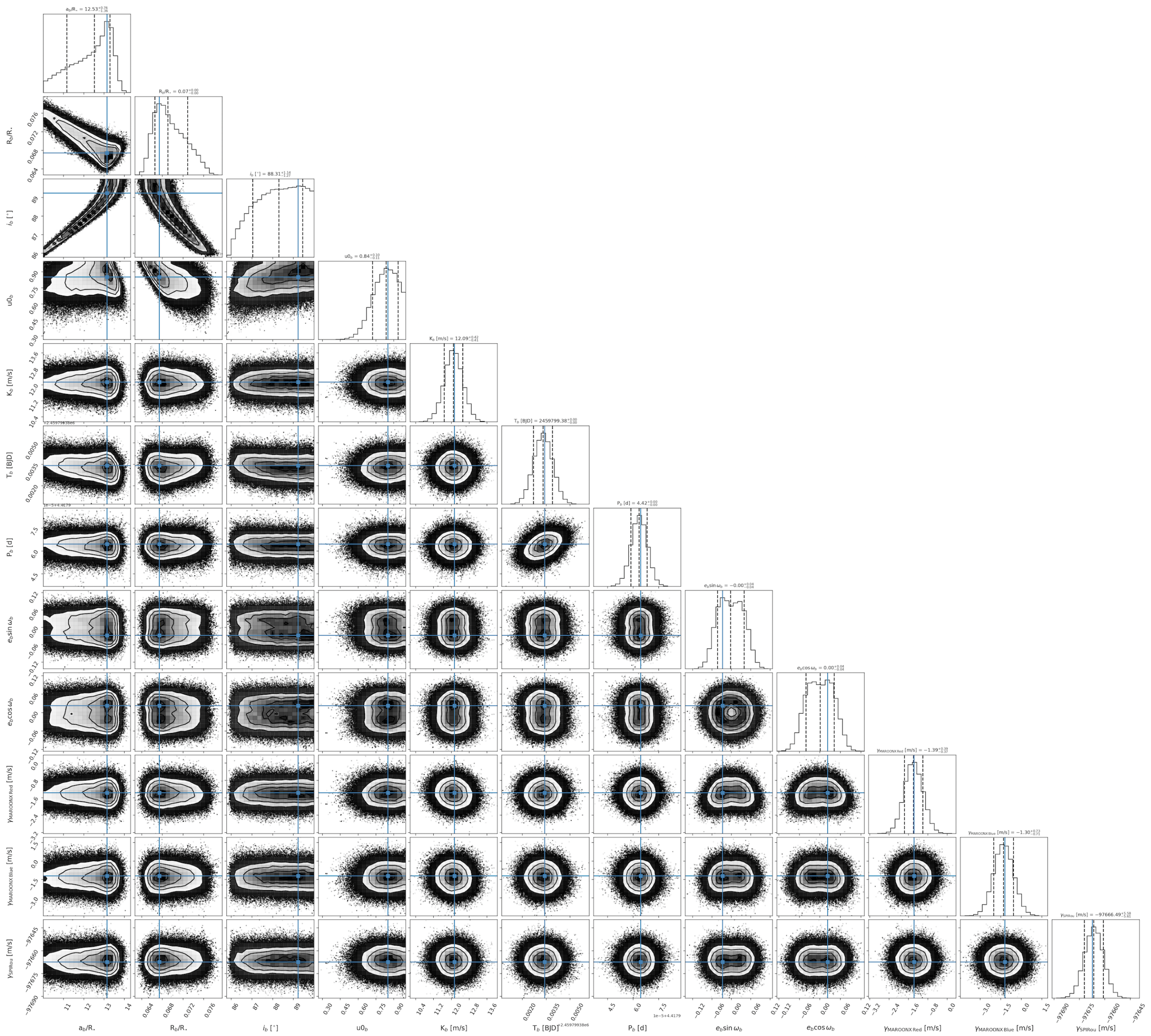}
\caption{Pairs plot showing the MCMC samples and posterior distributions of the free parameters in our joint analysis of the TESS and ground-based photometry and the MAROON-X and SPIRou RVs of TOI-3568.  The contours mark the 1$\sigma$, 2$\sigma$, and 3$\sigma$ regions of the distribution. The blue crosses indicate the best-fit values for each parameter obtained by the mode, and the dashed vertical lines in the projected distributions show the median values and the 1$\sigma$ uncertainty (34\% on each side of the median).}
\label{fig:TOI-3568_pairsplot}
\end{figure*}

This appendix presents, in Figure \ref{fig:TOI-3568_pairsplot}, the MCMC samples and the final posterior distributions of the free parameters used in our joint analysis of the TESS and ground-based photometry, as well as the MAROON-X and SPIRou RVs of TOI-3568.

\end{appendix}

\end{document}